\documentclass{article}

\usepackage[final]{neurips_2023}

\usepackage[utf8]{inputenc} %
\usepackage[T1]{fontenc}    %
\usepackage{hyperref}       %
\usepackage{url}            %
\usepackage{booktabs}       %
\usepackage{amsfonts}       %
\usepackage{nicefrac}       %
\usepackage{microtype}      %
\usepackage{xcolor}         %

\usepackage{amsmath}       %
\usepackage{enumerate}
\usepackage{tabularx}       %
\usepackage{array,multirow,graphicx}
\usepackage{algorithm,algorithmic}
\usepackage{mathtools}
\usepackage{cleveref} %
\crefname{figure}{figure}{figures}
\crefname{equation}{equation}{equations}
\crefname{table}{table}{tables}
\crefname{section}{section}{sections}
\usepackage{subcaption}
\usepackage[
    separate-uncertainty=true,
    detect-weight,
    table-number-alignment = center,
	table-text-alignment = center,
	table-format=3.2(4),
    table-sign-mantissa,
]{siunitx}
\usepackage{etoolbox}
\usepackage{todonotes}

\newcommand{\cN}{\mathcal{N}}
\newcommand{\cI}{\mathcal{I}}
\newcommand{\cO}{\mathcal{O}}
\newcommand{\jO}{\mathbf{\mathcal{O}}}
\newcommand{\cP}{\mathcal{P}}

\newcommand{\cS}{\mathcal{S}}
\newcommand{\cA}{\mathcal{A}}
\newcommand{\jA}{\mathbf{\mathcal{A}}}

\newcommand{\ttrain}{\mathcal{T}_{\text{train}}}
\newcommand{\ttest}{\mathcal{T}_{\text{test}}}
\newcommand{\Ntrain}{N_\text{train}}
\newcommand{\Ntest}{N_{\text{test}}}

\newcommand{\bpi}{\boldsymbol{\pi}}
\newcommand{\btau}{\boldsymbol{\tau}}
\newcommand{\bmu}{\boldsymbol{\mu}}
\newcommand{\bsigma}{\boldsymbol{\sigma}}
\newcommand{\bpsi}{\boldsymbol{\psi}}
\newcommand{\bphi}{\boldsymbol{\phi}}

\title{Learning Task Embeddings for Teamwork Adaptation in Multi-Agent Reinforcement Learning}

\author{%
  Lukas Sch\"afer\\
  University of Edinburgh\\
  \texttt{l.schaefer@ed.ac.uk} \\
  \And
  Filippos Christianos\\
  University of Edinburgh\\
  \texttt{f.christianos@ed.ac.uk} \\
  \AND
  Amos Storkey\\
  University of Edinburgh\\
  \texttt{a.storkey@ed.ac.uk} \\
  \And
  Stefano V. Albrecht\\
  University of Edinburgh\\
  \texttt{s.albrecht@ed.ac.uk} \\
}

\begin{document}

\maketitle

\begin{abstract}
    Successful deployment of multi-agent reinforcement learning often requires agents to adapt their behaviour. In this work, we discuss the problem of teamwork adaptation in which a team of agents needs to adapt their policies to solve novel tasks with limited fine-tuning. Motivated by the intuition that agents need to be able to identify and distinguish tasks in order to adapt their behaviour to the current task, we propose to learn \textit{multi-agent task embeddings} (MATE). These task embeddings are trained using an encoder-decoder architecture optimised for reconstruction of the transition and reward functions which uniquely identify tasks. We show that a team of agents is able to adapt to novel tasks when provided with task embeddings. We propose three MATE training paradigms: independent MATE, centralised MATE, and mixed MATE which vary in the information used for the task encoding. We show that the embeddings learned by MATE identify tasks and provide useful information which agents leverage during adaptation to novel tasks. %
\end{abstract}

\section{Introduction}
\label{sec:introduction}
\textit{Multi-agent reinforcement learning} (MARL) is a machine learning paradigm which enables multiple agents to concurrently learn behaviour from interactions with the environment as well as interactions with each other. MARL methods have become increasingly capable of learning complex behaviour~\citep{berner2019dota,vinyals2019grandmaster}, but their learned behaviours are usually highly task-specific. This can be desirable to maximise effectiveness in specific tasks, but limits the applicability in the real-world, which often requires the learned behaviour to be robust to small perturbations and changes in the environment~\citep{dulac2021challenges,akkaya2019solving}.

\begin{figure}
    \centering
    \includegraphics[width=.8\textwidth]{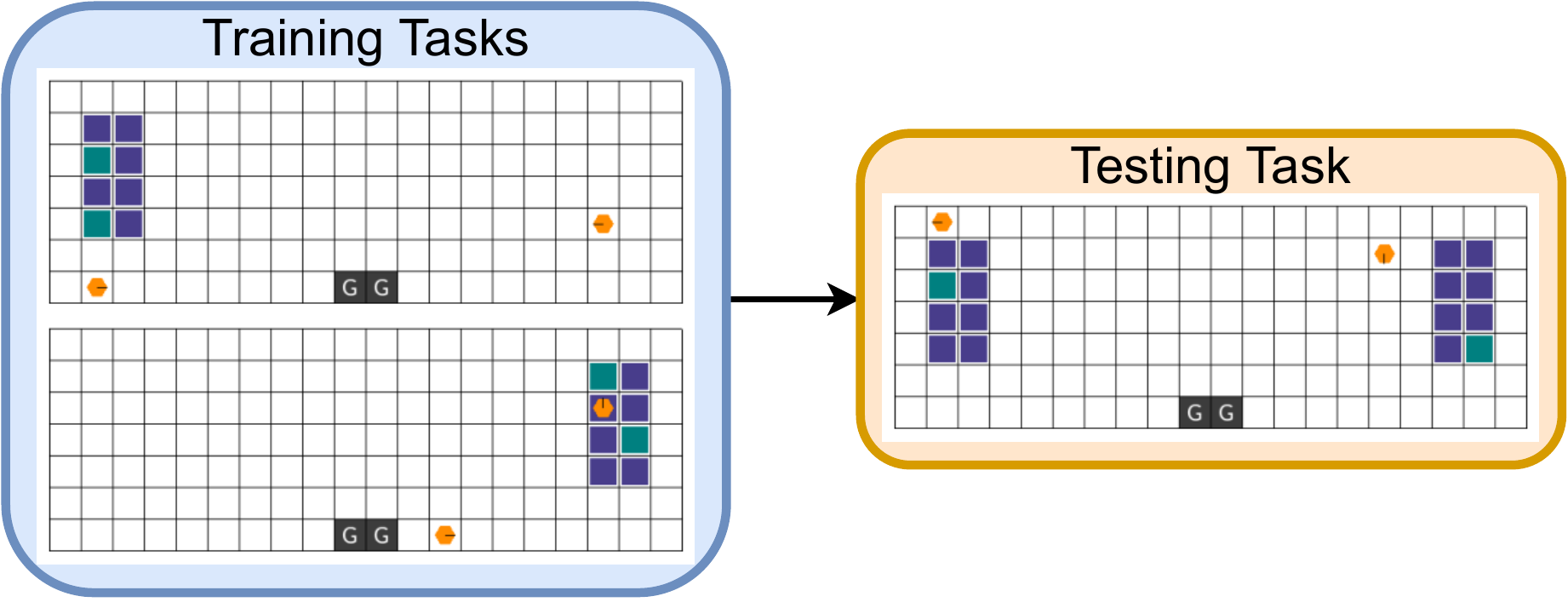}
    \caption{Teamwork adaptation in multi-robot warehouse environment: In the training tasks with different layouts, both agents (orange) need to deliver shelves (blue) with requested items (green) to the delivery zone (black). After training, the agents are fine-tuned and evaluated in the unseen testing task which requires novel coordination to effectively distribute the workload across both agents.}
    \label{fig:rware_generalisation}
\end{figure}

Our work addresses the challenge of \textbf{teamwork adaptation} in which a team of agents is trained in a set of training tasks, and then has to adapt to novel, previously unseen testing tasks.
As an example, consider a warehouse environment in which a team of agents must navigate in the warehouse to collect and deliver shelves with requested items (\Cref{fig:rware_generalisation}). Tasks can vary in their warehouse layout, and agents need to adjust their team strategy depending on the layout of the warehouse to optimise deliveries. Whereas the warehouses with shelves on one side (\Cref{fig:rware_generalisation}, left) require all agents to move to the same side of the warehouse, the warehouse with shelves on both ends (\Cref{fig:rware_generalisation}, right) requires agents to effectively split between both ends to minimise waiting and travel time.

The above example illustrates that a team of agents may need to adapt their individual behaviour and coordination to novel tasks in a non-trivial way. However, such adaptation requires agents to identify the task they are in. Current MARL approaches are not designed for such adaptation, and in our experiments we show their limitations at adapting to novel tasks. %

Motivated by the challenge of teamwork adaptation, we propose to equip agents with the ability to infer the task they are in by learning \textbf{multi-agent task embeddings} (MATE). Through interacting with the environment, each agent builds up a task embedding using an encoder conditioned on the trajectories of agents in the task. 
Motivated by the observation that each task is uniquely identifiable by its transition and reward functions, the encoder is jointly trained with a separate decoder which conditioned on task embeddings is optimised to reconstruct transitions and rewards. %
Learned embeddings allow a team of agents to adapt to previously unseen but related tasks by conditioning their policies on task embeddings. We hypothesise that access to task embeddings simplifies teamwork adaptation. %
We consider and compare three paradigms of learning such embeddings: independent MATE, centralised MATE, and mixed MATE. For independent MATE, each agent independently encodes local information to obtain an embedding of the task, whereas centralised MATE trains a single encoder conditioned on the joint information for all agents. Mixed MATE trains independent encoders for agents but optimises them using a mixture of task embeddings with weights for each task embedding being conditioned on the state of the environment.

We empirically evaluate all MATE variants in four multi-agent environments which require increasing degrees of adaptation. We find that MATE improves teamwork adaptation in several cases compared to fine-tuning MARL agents without task embeddings or training agents in testing tasks with no prior training. In particular, in tasks where all agents need to coordinate, MATE provides useful information for agent adaptation, with mixed MATE performing the best among all three paradigms across a wide range of adaptation scenarios. We further demonstrate that learned task embeddings produce clusters which clearly identify tasks, and the learned mixture of mixed MATE focuses on task embeddings of agents which currently observe the most information about the task.

\section{Related Work}
\label{sec:related_work}

\textbf{Meta Reinforcement Learning:} Meta reinforcement learning (Meta RL) aims to leverage concepts from meta learning~\citep{hospedales2020meta} to learn policies which can adapt to novel testing tasks using limited interactions. Such adaptation can be achieved by computing meta gradients to find a model initialisation from which an effective policy in testing tasks can be obtained using few optimisation steps, e.g.\ using MAML~\citep{finn2017model} or REPTILE~\citep{nichol2018first}. %
Another common approach for meta RL implicitly builds up a latent context using recurrent neural networks which enable adaptation to new tasks~\citep{duan2016rl,wang2016learning,fakoor2020meta}. %
However, these approaches rely on the recurrent context to learn useful information for adaptation without any explicit optimisation enforcing such usefulness. \citet{rakelly2019efficient} disentangle the objectives of task inference and learning control policies and leverage variational inference techniques to learn a context inference model. \citet{zintgraf2020varibad} also propose to learn a variational inference model to explicitly learn task beliefs and incorporate task uncertainty into the policies for adaptation. Their approach is similar in architecture and optimisation to MATE but is limited to the adaptation of policies of individual agents. In contrast, we address the challenge of adapting the policies of multiple agents cooperating in a team.

\textbf{Transfer Learning:} Established techniques from transfer learning~\citep{taylor2009transfer} address a similar challenge to meta RL by extracting representations, action selection or other components from already learned models to improve the capabilities of agents in novel testing tasks. Unlike meta RL which typically adapts to novel tasks using fine-tuning over few episodes, transfer learning requires a dedicated transferring procedure for each new task. \citet{da2019survey} provide an overview specifically for methods aiming to transfer policies of multiple agents. They distinguish between inter-agent transfer, aiming to leverage information from one agent with possibly more expertise to transfer another agent~\citep{da2017simultaneously,da2018autonomously}, and intra-agent transfer which aims to transfer knowledge across tasks for all involved agents. Related to transfer learning are approaches focusing on multi-task learning in which agents are trained to generalise over a set of multiple tasks~\citep{vithayathilvargheseSurveyMultiTaskDeep2020,omidshafiei2017deep,caruana1997multitask}.%

\looseness=-1
\textbf{Multi-Agent Reinforcement Learning:} There has been significant progress in MARL to solve challenging coordination problems~\citep{papoudakis2021benchmarking} with many approaches focusing on the paradigm of centralised training and decentralised execution (CTDE)~\citep{christianos2020shared,rashid2018qmix,foerster2018counterfactual,sunehag2017value,lowe2017multi}. Under this paradigm, information is shared across agents at training time without conditioning agent policies on such joint information. In this way, training can leverage privileged information without preventing agent policies from being deployed in a decentralised manner. 
We train MATE using the reconstruction of transition and reward functions of tasks defined over all agents, so MATE also falls under the paradigm of CTDE. In multi-agent systems, it is appealing to explicitly learn models of the behaviour of other agents in the environment to infer their possible behaviour~\citep{papoudakis2021agent,zintgraf2021deep,albrecht2018autonomous}. Such approaches demonstrate the ability for a single agent within a multi-agent system to adapt to different agents to cooperate or compete with. Similarly, the challenges of ad hoc teamwork~\citep{stone2010ad} and zero-shot coordination~\citep{hu2020other} address the problem of training agents to be able to coordinate with new partners without prior training in a team or established team strategies. Approaches in these settings prominently model other agents~\citep{rahman2021towards,barrett2015cooperating} or learn policies to effectively play alongside a diverse population of agents maintained during training~\citep{lupu2021trajectory,qiu2023rpm}. However, few work exist which address the challenge of adapting the policies of multiple agents to novel tasks. \citet{hu2021updet} propose a novel architecture leveraging transformer models and self-attention~\citep{vaswani2017attention} to be able to reuse policies for varying team sizes, and \citet{zhang2021learning} train a latent model to encode coordination information to generalise to varying team sizes. \citet{vezhnevetsOptionsResponsesGrounding2020} train agents using a hierarchical approach with a high-level policy choosing the low-level behaviour which should be deployed given currently available information about the task and other agents in the environment. \citet{liang2022continuous} apply the idea of REPTILE~\citep{nichol2018first} to MARL, finding an effective initialisation from training agent policies on multiple training tasks which is shown to generalise to new testing tasks. \citet{mahajan2022generalization} formalise the problem of combinatorial generalisation in which a team of agents needs to generalise over varying capabilities of agents. In contrast to all these approaches, we focus on the problem of adapting a fixed team of agents to varying tasks.

\section{Problem Definition}
\label{sec:problem_setting}
\paragraph{Partially-Observable Stochastic Games} We consider multi-agent tasks modelled as partially observable stochastic games (POSGs) for $N$ agents~\citep{hansen2004dynamic}. %
A POSG is given by the tuple $(\cI,\cS, \{\cO^i\}_{i\in \cI}, \{A^i\}_{i\in \cI}, \Omega, \cP, \{R^i\}_{i\in \cI})$. 
Agents are indexed by $i\in\cI = \{1,\ldots,N\}$. $\cS$ denotes the state space of the environment and $\jA = A^1\times\ldots\times A^N$ denotes its joint action space. $\cP: \cS \times \jA \times \cS \mapsto [0, 1]$ denotes the transition function of the environment, defining a distribution of successor states given the current state and the applied joint action. At each time step $t$, each agent $i$ receives an observation $o_t^i \in \cO^i$ defined by the observation function $\Omega: \cS \times \jA \mapsto \Delta(\jO)$ conditioned on the current state and applied joint action for joint observation space $\jO = \cO^1\times\ldots\times \cO^N$. Each agent learns a policy $\pi_i(a^i_t | o^i_{1:t})$ conditioned on its history of observations, denoted by $o^i_{1:t} = (o^i_1, \ldots, o^i_t)$. After time step $t$, agent $i$ receives a reward $r^i_t$ given by its reward function $R^i: \cS \times \cA \mapsto \mathbb{R}$. The objective is to learn a joint policy $\bpi = (\pi_1, \ldots, \pi_N)$ such that the discounted returns of each agent $G^i = \sum_{t=1}^\infty \gamma^{t-1} r^i_t$ are maximised with respect to the policies of all other agents, formally $\forall_{i}: \pi_{i} \in \arg \max _{\pi_{i}^{\prime}} \mathbb{E}\left[G^i | \pi_{i}^{\prime}, \pi_{-i}\right]$, with discount factor $\gamma \in [0; 1)$ and $\pi_{-i} = \bpi \setminus \{\pi_i\}$.

\paragraph{Teamwork Adaptation}
\looseness=-1
We consider the challenge of transferring the joint policies $\bpi$ of a fixed team of agents from a set of training tasks $\ttrain$ to novel testing tasks $\ttest$ with $\ttrain \cap \ttest = \emptyset$. Each task is represented as a POSG. %
Without any further assumptions, training and testing tasks could be arbitrarily different and hence no generalisation could be feasibly expected. %
We assume that training and testing tasks have identical number of agents $N$, action space $\cA$ and identical dimensionality of observations. 
In our problem setting, agents are first trained in $\ttrain$ for $\Ntrain$ time steps and can be trained in $\ttest$ for a limited number of time steps $\Ntest$ to fine-tune the policy. We refer to this setting as teamwork adaptation which is similar to the problem of domain adaptation for RL~\citep{eysenbach2020off}.

\section{Learning Multi-Agent Task Embeddings}

\begin{figure}[t]
    \centering
    \includegraphics[width=\linewidth]{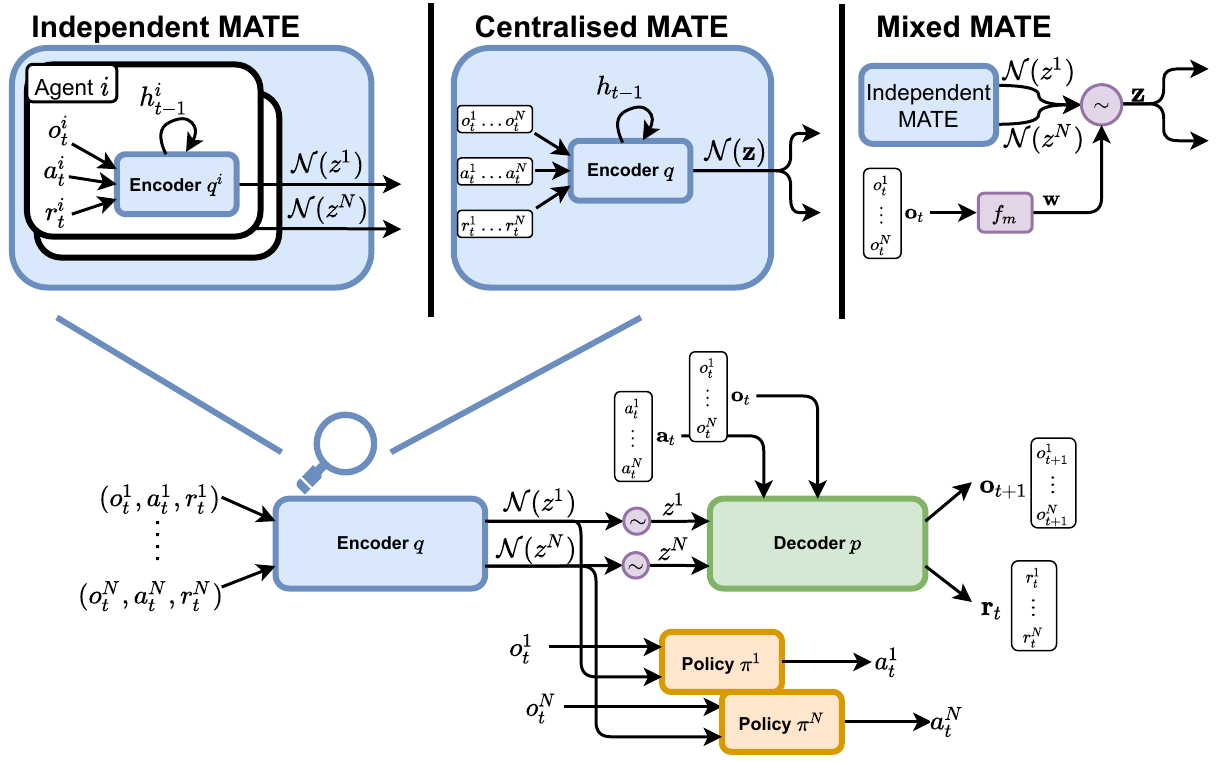}
    \caption{MATE recurrently encodes (blue) trajectories of states, actions and rewards into variational task embeddings. We consider three paradigms of MATE: independent MATE, centralised MATE and mixed MATE. Policies (orange) are conditioned on task embeddings and the decoder (green) receives sampled (purple) task embeddings, current observations and actions to predict the transition and reward function of all agents.}
    \label{fig:mate}
\end{figure}

For agents to be able to adapt their behaviour to the current task, they need to infer the task they are in as they interact with their environment. Therefore, we propose to learn multi-agent task embeddings (MATE) $\cN(\mathbf{z})$ from agent trajectories, consisting of observations, actions and rewards, which provide agent policies $\pi_i(a^i_t | o^i_{1:t}, \cN(\mathbf{z}))$ with information about the task as well as a measure of uncertainty about the task. We hypothesise that explicitly modelling and learning task embeddings is advantageous for teamwork adaptation. %

In order to learn task embeddings, we train a variational encoder-decoder architecture~\citep{kingma2013auto}. Agents use an encoder $q$ to compute a variational distribution $q(\mathbf{z} | \btau_{1:t}) = \cN(\mathbf{z}; \bmu, \text{diag}(\bsigma))$ representing the task embeddings. These task embeddings are given by a diagonal Gaussian distribution with mean $\bmu \in \mathbb{R}^d$ and variance $\bsigma \in \mathbb{R}_+^d$. We concisely denote this distribution over task embeddings as $\cN(\mathbf{z})$. The encoder is represented as a recurrent neural network which outputs $\bmu$ and $\log\left(\text{diag}(\bsigma)\right)$. At each time step, information about the trajectory consisting of observations, actions and rewards is fed into the encoder together with the previous hidden state to update the task embeddings. The hidden state is reset after each episode to enable training on multiple tasks sampled for each episode. In order to train the encoder, we define a decoder $p$ to reconstruct the transition and reward functions across all tasks conditioned on task embedding samples $\mathbf{z} \sim \cN(\mathbf{z})$.
The encoder and decoder, parameterised by $\bphi$ and $\bpsi$ respectively, are jointly optimised to maximise the following evidence lower bound (ELBO) given trajectory $\btau_{1:t}$
\begin{equation}
    \text{ELBO}(\bphi, \bpsi | \btau_{1:t}) = \mathbb{E}_{q\left({\mathbf{z}_t | \btau_{1:t}; \bphi}\right)}\left[\log p({\mathbf{o}_{t+1}, \mathbf{r}_t | \mathbf{o}_t, \mathbf{a}_t, \mathbf{z}; \bpsi})\right] - \beta \text{KL}\left(q\left({\mathbf{z} | \btau_{1:t}; \bphi}\right)|| p(\mathbf{z})\right)
    \label{eq:vae_elbo}
\end{equation}
with additional hyperparameter $\beta$ to control the regularisation of the KL prior~\citep{higgins2017beta}. This objective is motivated by the observation that each task can be identified by its unique transition and reward functions. Modelling the decoder as a multivariate Gaussian model over observations and rewards with constant diagonal covariance matrix and assuming a standard Gaussian prior allows us to minimise the following loss, equivalent to maximising the ELBO (see \Cref{app:mate_loss_derivation} for derivation):
\begin{multline}
    \mathbb{L}(\bphi, \bpsi | \btau_{1:t})  = \mathbb{E}_{q\left({\mathbf{z}_t | \btau_{1:t}; \bphi}\right)}\Big[\left(p({\mathbf{o}_{t+1} | \mathbf{o}_t, \mathbf{a}_t, \mathbf{z}; \bpsi}) - \mathbf{o}_{t+1}\right)^2 \\+ \left(p({\mathbf{r}_t | \mathbf{o}_t, \mathbf{a}_t, \mathbf{z}; \bpsi}) - \mathbf{r}_t\right)^2\Big] - \beta \frac{1}{2} \sum_{j=1}^{d}\left(1+\log \left(\sigma_{j}^{2}\right)-\mu_{j}^{2}-\sigma_{j}^{2}\right)
    \label{eq:mate_loss}
\end{multline}
The intuition of this loss is that the decoder $p$ is optimised to model the mean of the generative multivariate Gaussian distribution over reconstructed observations and rewards. %
We consider three paradigms of learning MATE: (1) independent MATE, (2) centralised MATE and (3) mixed MATE which vary in the information used to encode task embeddings.

\textbf{Independent Multi-Agent Task Embeddings (Ind-MATE)} independently trains separate encoders $q^i(z^i | \tau^i_{1:t}; \phi^i)$ with $\tau^i_{1:t} = \{(o^i_u, a^i_u, r^i_u)\}_{u=1}^t$ for each agent $i$ conditioned only on its individual trajectory. The centralised decoder is shared across all agents and used to decode individual task embeddings of all agents, and the policy $\pi_i(a_t^i | o_t^i, \cN(z^i))$ of agent $i$ is conditioned only on its individual task embedding. We hypothesise that such task embeddings are limited in the encoded information which cannot fully represent the task under partial observability.

\textbf{Centralised Multi-Agent Task Embeddings (Cen-MATE)} instead shares a single encoder $q(\mathbf{z} | \btau_{1:t}; \bphi)$ with $\btau_{1:t} = \{(\mathbf{o}_u, \mathbf{a}_u, \mathbf{r}_u)\}_{u=1}^t$ and decoder conditioned on the joint information across all agents. The policy of each agent $i$ is conditioned on the joint, shared task embedding $\cN(\mathbf{z})$. Such a shared, centralised task embedding has access to information from all agents and can therefore encode more information about the task than Ind-MATE. However, policies depend on the computation of the task embedding which requires access to the joint information across all agents. The access to this privileged information prevents decentralised execution of agents.

\looseness=-1
\textbf{Mixed Multi-Agent Task Embeddings (Mix-MATE)} is similar to Ind-MATE in that each agent trains an individual encoder conditioned only on its local trajectory to ensure decentralised execution. However, instead of decoding each variational task embedding independently, a mixture of task embeddings is computed $q(\mathbf{z} | \btau_{1:t}; \bphi) = \sum_{i=1}^N w_i q^i(z^i | \tau^i_{1:t}; \phi^i)$ and sampled from. Mixture weights $\mathbf{w} = f_m(\mathbf{o}_t)$ are computed using a single-layer network with softmax output conditioned on the joint observations. %
We hypothesise that such mixing allows individual task embeddings to be more representative of the full task while preserving decentralised execution. We also note that the mixture distribution provides insight into which agent's task embedding is considered most important in a given state.

\section{Experimental Evaluation}
\label{sec:evaluation}
\begin{figure}
    \centering
    \begin{subfigure}{.24\textwidth}
        \centering
        \includegraphics[height=10em]{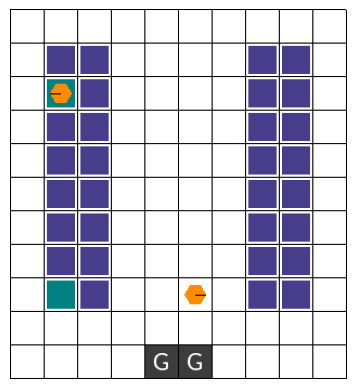}
        \caption{\centering Multi-robot warehouse (RWARE)}
    \end{subfigure}
    \begin{subfigure}{.24\textwidth}
        \centering
        \fbox{\includegraphics[height=8.9em]{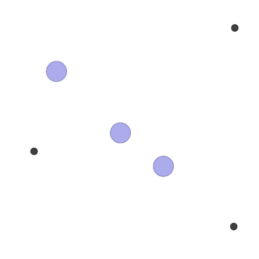}}
        \caption{\centering Multi-agent particle (MPE)}
    \end{subfigure}
    \begin{subfigure}{.24\textwidth}
        \centering
        \includegraphics[height=10em]{media/tasks/bpush-small-2ag}
        \caption{\centering Boulder-push (BPUSH)}
    \end{subfigure}
    \begin{subfigure}{.24\textwidth}
        \centering
        \includegraphics[height=10em]{media/tasks/lbf-grid-8x8-2p-2f}
        \caption{\centering Level-based foraging (LBF)}
    \end{subfigure}
    \caption{Visualisations of all four multi-agent environments}
    \label{fig:envs}
\end{figure}

In this section, we evaluate our proposed approach of learning multi-agent task embeddings. In particular, we will investigate (1) whether MATE improves teamwork adaptation given by returns in testing tasks after limited fine-tuning, (2) how the three paradigms of MATE compare to each other, and (3) what information is encoded by MATE. To answer these questions, we conduct an evaluation in four multi-agent environments, visualised in \Cref{fig:envs}. In all experiments, we train agents using the multi-agent synchronous advantage actor-critic (MAA2C)~\citep{papoudakis2021benchmarking,mnih2016asynchronous} algorithm with recurrent policies containing gated recurrent units (GRU)~\citep{cho2014learning}. During fine-tuning, we freeze MATE encoders and only fine-tune policies and critics. For further implementation details, see \Cref{app:imp_details}, and details for each task can be found in \Cref{app:marl_envs}.

\subsection{Multi-Agent Environments}
\label{sec:evaluation_envs}

\textbf{Multi-Robot Warehouse:} In our motivational example of the multi-robot warehouse (RWARE)~\citep{papoudakis2021benchmarking,christianos2020shared} agents need to collect and deliver requested items from shelves within warehouses. Each agent observes a $5\times5$ grid centred around the agent containing information about nearby shelves, agents and delivery zones. Agents can move forward, pick-up shelves, and rotate which also rotates their observation. Tasks include warehouses with two or four agents, and varying layouts which lead to static, but significant variation requiring adaptation in coordination behaviour outlined in \Cref{fig:rware_generalisation}. Training and testing sets include multiple warehouses of similar layout but a range of sizes. Each episode consists of $500$ time steps.

\looseness=-1
\textbf{Multi-Agent Particle Environment:} In the cooperative navigation task of the multi-agent particle environment (MPE)~\citep{mordatch2018emergence,lowe2017multi}, three agents need to navigate a continuous two-dimensional, fully-observable world to cover all three landmarks which are initialised in random locations while avoiding collisions with each other. In the training task, agents only receive a small punishment of $-1$ for collisions with each other whereas testing tasks punish agents with significantly higher negative rewards of $-5$ and $-50$, respectively. Episodes terminate after $25$ time steps.

\looseness=-1
\textbf{Boulder-Push:} In this new environment (BPUSH), agents navigate a gridworld and need to push a box towards a goal location. Agents observe the box and agents in a $9\times9$ grid and the direction the box is required to be pushed in. These tasks require significant coordination with agents only receiving rewards for successfully pushing the box forward which requires cooperation of all agents. Episodes terminate after the box has been pushed to its goal location or after at most $50$ time steps. In the training task, two agents need to push a box in a small $8\times8$ gridworld. Testing tasks include tasks with gridworld sizes of $12\times12$ (medium) and $20\times20$ (large) as well as tasks with a small negative penalty of $-0.01$ for unsuccessful pushing attempts of agents. Tasks with penalty terms make the exploration of the optimal behaviour of only pushing the box together with the other agent significantly harder.

\looseness=-1
\textbf{Level-Based Foraging:} In the level-based foraging (LBF) environment~\citep{Albrecht2013ASystems,albrecht2018autonomous}, multiple agents need to coordinate in a gridworld to pick-up food. Food and agents are assigned levels with agents only being able to pick-up adjacent food if the sum of levels of all agents cooperating to pick-up is greater or equal to the level of the food. Agents observe a $5\times5$ grid centred on themselves and are rewarded for picking-up food depending on its level and their contribution. Episodes terminate after all food has been collected or after at most $50$ time steps. We train in two comparably simple LBF tasks with gridworld sizes of $8\times8$ and two and four agents, respectively. Testing tasks for two and four agents contain a task with gridworld size $10\times10$ in which each food will require at least two agents to coordinate to pick it up, tasks with larger gridworld size of $15\times15$ and tasks with a small penalty of $-0.1$ for unsuccessful picking attempts of food.

\subsection{Baselines}
\label{sec:evaluation_baselines}
\textbf{No MATE (fine-tune)}: For the fine-tune baseline, we train MARL agents without any task embeddings in the training task and fine-tune in testing tasks for further $\Ntest$ time steps. This baseline allows us to evaluate the benefits of task embeddings for teamwork adaptation.

\textbf{No MATE (scratch)}: For the scratch baseline, we train MARL agents without any task embeddings only in the testing tasks for $\Ntest$ time steps. This baseline allows us to distinguish adaptation settings where prior training in potentially simpler tasks can improve returns in testing tasks from cases of negative transfer where prior training hurts the performance of agents in the testing tasks.

We further highlight that the policies of all agents, including both baselines, contain GRU networks such that an implicit task embedding can be learned through the hidden state of the recurrent network.

\subsection{Evaluation Results}
\label{sec:evaluation_results}
\begin{table}[t]
    \caption{Fine-tuning performance given by IQM and standard deviation across final returns over five random seeds. Highest IQM per testing task (within one standard deviation) are shown in bold.}
    \label{tab:finetuning_results}
    \robustify\bf
    \resizebox{\textwidth}{!}{
        \begin{tabular}{@{}p{3em} l l S S S S S @{}}
            \toprule
            & $\ttrain$ ($\Ntrain$)     & $\ttest$ ($\Ntest$) & {Scratch}	& {Fine-tune}	& {Ind-MATE}	& {Cen-MATE}	& {Mix-MATE}\\ \midrule
            \multirow{6}{*}{RWARE} & \multirow{2}{*}{tiny-2ag (25M)} & small-2ag (25M) & 0.02(0)& \bf 8.54(229)& 5.97(341)& \bf 10.90(297)& \bf 7.96(411)\\
                 & & corridor-2ag (25M) & \bf 23.00(291)& \bf 22.26(558)& \bf 26.98(406)& \bf 28.59(710)& \bf 25.39(1017)\\
                 & \multirow{2}{*}{tiny-4ag (25M)} & small-4ag (25M) & 0.12(87)& \bf 28.48(92)& 26.61(114)& 26.58(132)& \bf 27.69(121)\\
                 & & corridor-4ag (25M) & \bf 50.13(184)& 42.75(370)& 43.12(231)& 45.53(242)& 43.96(330)\\
                 & wide-both (50M) & wide-one-sided (50M) & \bf 12.39(675)& 0.03(349)& 0.02(0)& 3.08(450)& 0.02(0)\\
                 & wide-one-sided (50M) & wide-both (50M) & 0.04(62)& \bf 7.06(395)& \bf 5.26(369)& 3.45(311)& \bf 7.34(339)\\\midrule
            \multirow{2}{*}{MPE} & \multirow{2}{*}{cooperative navigation (10M)} & penalty navigation, pen=5 (10M) & \bf -259.65(1109)& -258.81(113)& \bf -257.50(31)& -258.47(75)& -258.45(68)\\
                 & & penalty navigation, pen=50 (10M) & -2019.35(395)& \bf -2016.83(354)& -2019.20(379)& -2020.60(439)& \bf -2015.15(201)\\\midrule
            \multirow{4}{*}{BPUSH} & \multirow{4}{*}{small-2ag (5M)} & small-pen-2ag (20M) & \bf -0.01(1)& \bf -0.00(105)& \bf 0.88(130)& \bf -0.00(103)& \bf 0.89(130)\\
                 & & medium-pen-2ag (20M) & \bf -0.01(1)& \bf -0.00(0)& \bf -0.00(0)& \bf 0.00(0)& \bf 0.00(0)\\
                 & & medium-2ag (20M) & 0.69(36)& 1.24(47)& \bf 2.04(29)& \bf 2.07(46)& \bf 2.32(52)\\
                 & & large-2ag (20M) & 0.00(0)& 0.39(10)& \bf 0.60(16)& 0.43(14)& \bf 0.97(40)\\\midrule
            \multirow{6}{*}{LBF} & \multirow{3}{*}{8x8-2p-2f (5M)} & 10x10-2p-2f-coop (5M) & \bf 0.88(2)& 0.80(3)& \bf 0.84(4)& 0.81(3)& \bf 0.84(4)\\
                 & & 15x15-2p-4f (5M) & \bf 0.71(3)& \bf 0.67(4)& \bf 0.73(8)& \bf 0.71(1)& \bf 0.71(2)\\
                 & & 15x15-2p-2f-pen (5M) & \bf 0.70(0)& \bf 0.64(8)& 0.55(6)& 0.65(4)& 0.56(12)\\
                 & \multirow{3}{*}{8x8-4p-4f (5M)} & 10x10-4p-2f-coop (5M) & \bf 0.78(1)& \bf 0.73(5)& 0.74(3)& 0.72(2)& 0.74(2)\\
                 & & 15x15-4p-6f (5M) & \bf 0.73(0)& 0.66(3)& \bf 0.71(2)& 0.69(2)& \bf 0.71(3)\\
                 & & 15x15-4p-4f-pen (5M) & \bf 0.44(0)& 0.30(8)& \bf 0.24(24)& \bf 0.35(10)& \bf 0.39(7)\\\bottomrule
        \end{tabular}
    }
\end{table}

\begin{figure}[t]
    \begin{subfigure}{\textwidth}
        \centering
        \includegraphics[trim={0 0.5em 0 0.5em},clip,width=.8\textwidth]{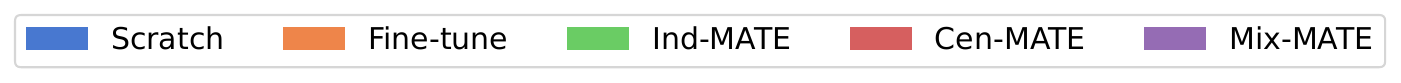}
    \end{subfigure}

    \begin{subfigure}{.33\textwidth}
        \centering
        \includegraphics[width=\textwidth]{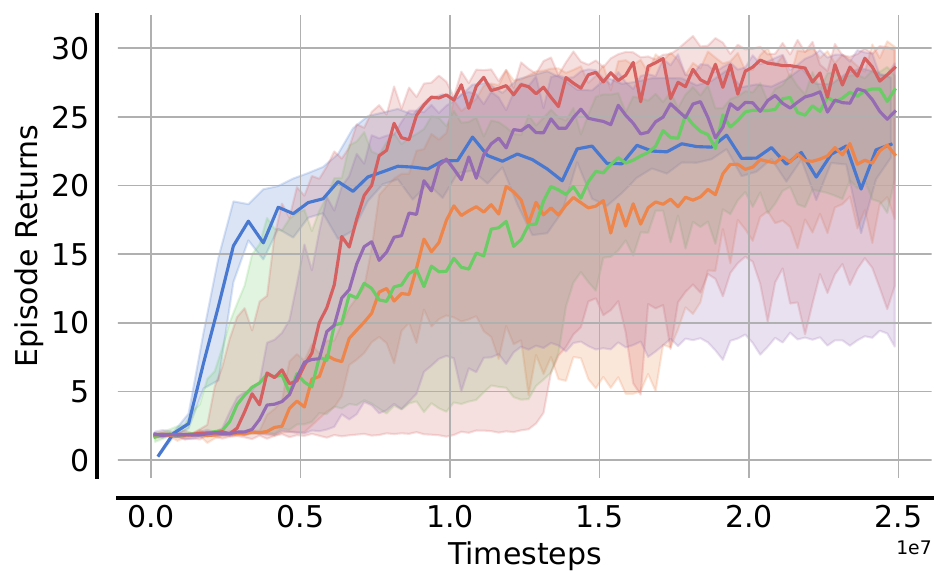}
        \caption{RWARE corridor-2ag}
        \label{fig:rware_corridor_2ag}
    \end{subfigure}
    \begin{subfigure}{.33\textwidth}
        \centering
        \includegraphics[width=\textwidth]{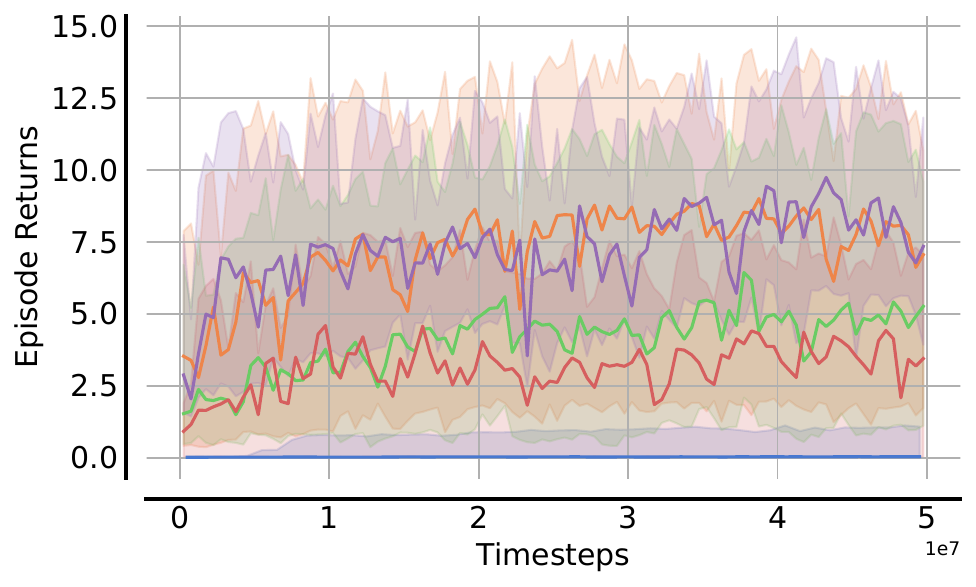}
        \caption{RWARE wide-both}
        \label{fig:rware_wide_both}
    \end{subfigure}
    \begin{subfigure}{.33\textwidth}
        \centering
        \includegraphics[width=\textwidth]{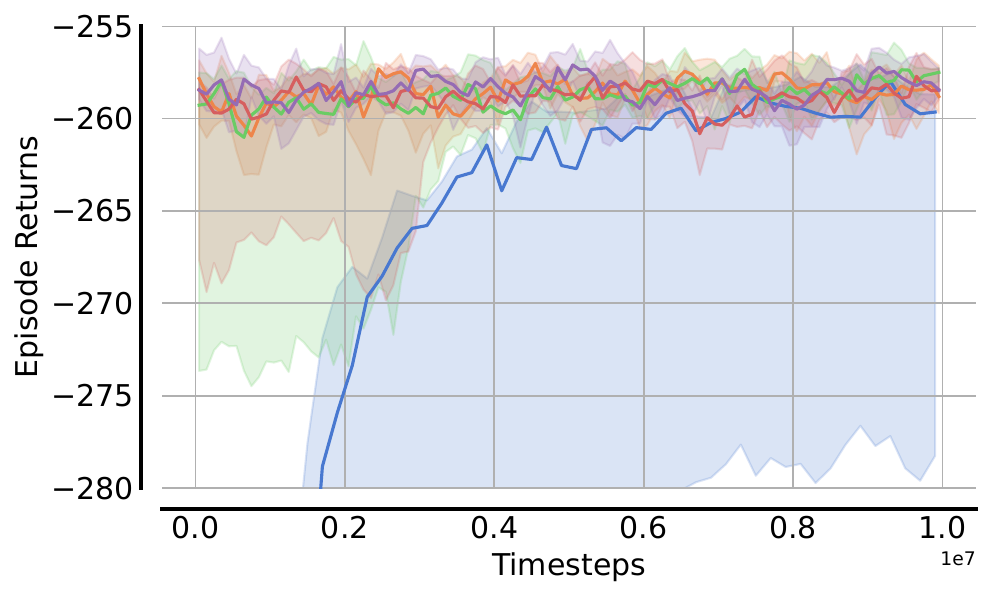}
        \caption{MPE penalty navigation (pen=5)}
        \label{fig:mpe_penalty_5}
    \end{subfigure}
    \begin{subfigure}{.33\textwidth}
        \centering
        \includegraphics[width=\textwidth]{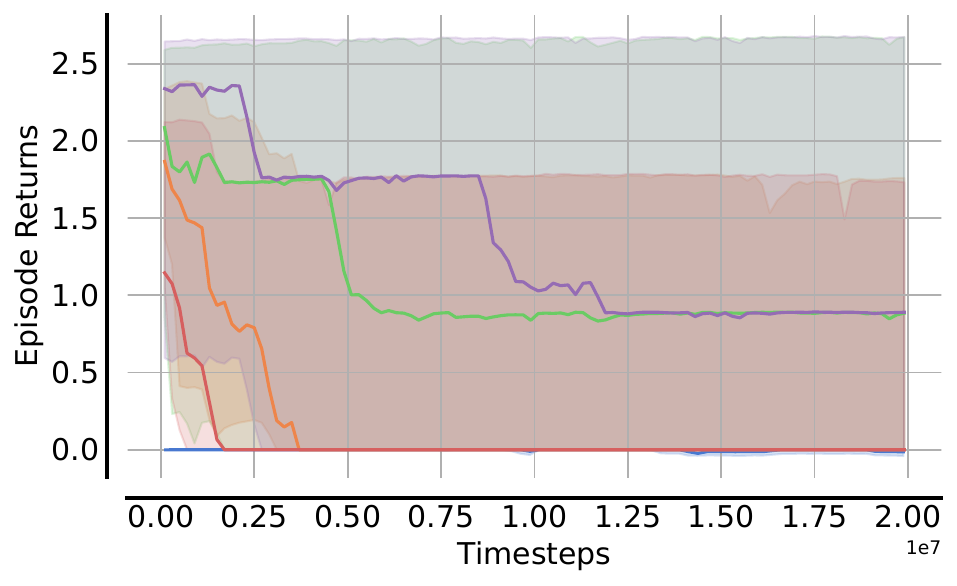}
        \caption{BPUSH small-pen-2ag}
        \label{fig:bpush_small_pen}
    \end{subfigure}
    \begin{subfigure}{.33\textwidth}
        \centering
        \includegraphics[width=\textwidth]{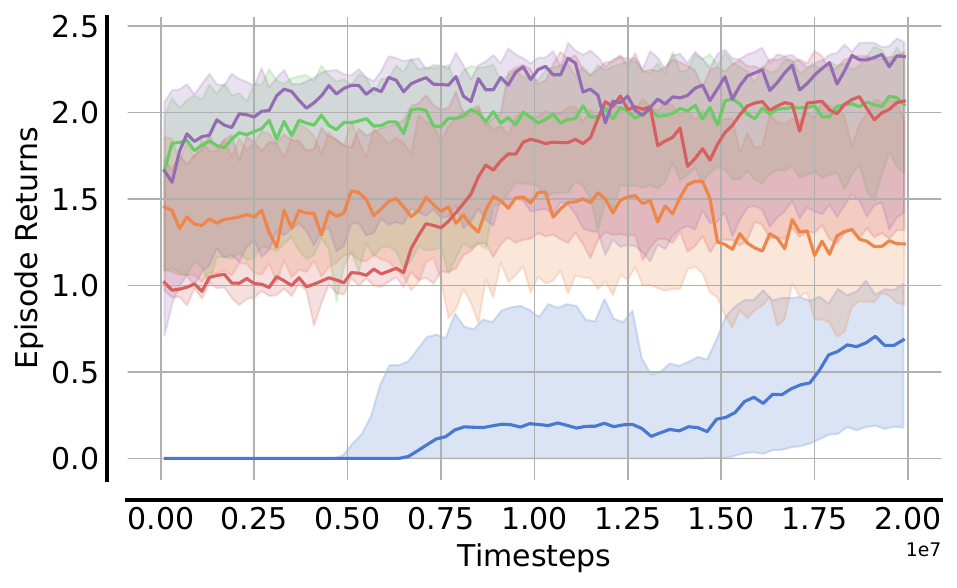}
        \caption{BPUSH medium-2ag}
        \label{fig:bpush_medium}
    \end{subfigure}
    \begin{subfigure}{.33\textwidth}
        \centering
        \includegraphics[width=\textwidth]{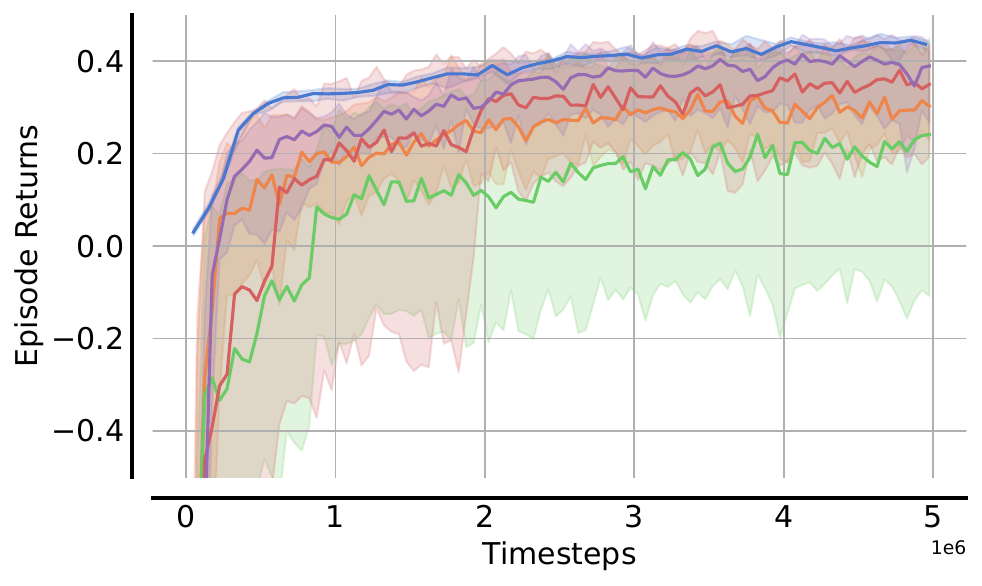}
        \caption{LBF 15x15-4p-4f-pen}
        \label{fig:lbf_4p_pen}
    \end{subfigure}
    \caption{IQM and 95\% confidence intervals for fine-tuning returns in selected testing tasks.}
    \label{fig:finetuning_results}
\end{figure}

We report episodic returns as the sum of episodic returns of all agents. We train and fine-tune each algorithm for five random seeds and report the interquartile mean (IQM)~\citep{agarwal2021deep} and standard deviation across final fine-tuning returns over all seeds in \Cref{tab:finetuning_results}. Learning curves with IQM and stratified bootstrap 95\% confidence intervals computed over 100,000 samples are shown in \Cref{fig:finetuning_results} for some testing tasks. Learning curves for all training and testing scenarios can be found in \Cref{app:training_plots,app:finetuning_plots}.

In RWARE (\Cref{fig:finetuning_rware_all}), agents need to heavily explore in order to receive positive rewards as they need to execute long sequences of actions to deliver requested items. Therefore, agents significantly benefit from the initial training phase during fine-tuning for testing tasks with larger or more complicated environments. In particular, in small warehouse instances and wide warehouses with shelves on both sides (\Cref{fig:rware_wide_both}) we find the fine-tuning of agents with and without MATE improves returns significantly compared to training from scratch. In contrast, in simpler environments such as the corridor warehouses (\Cref{fig:rware_corridor_2ag}) agents appear to suffer from the initial training phase starting with policies which exhibit less exploration than randomly initialised policies with entropy regularisation. For MATE, we find that Cen-MATE with access to observations of all agents performs the best in most tasks due to significant partial observability in these tasks.

Fine-tuning agents in MPE (\Cref{fig:finetuning_mpe_all}) from simple spread to tasks with larger collision penalties appears to lead to no significant benefits over training agents from scratch. Agent policies remain almost identical throughout fine-tuning for both testing tasks, exhibiting almost identical returns at the beginning and end of fine-tuning.%

For BPUSH (\Cref{fig:finetuning_bpush_all}), we find MATE significantly outperforms all baselines in three out of four fine-tuning scenarios. In particular Mix-MATE outperforms all other approaches. Adding a penalty to unsuccessful pushing attempts of agents makes the exploration of the optimal behaviour significantly more challenging, so training agents from scratch is unsuccessful. Agents previously trained on the simpler task without such a penalty perform well early during fine-tuning, but throughout fine-tuning agents exhibit more defensive behaviour and refrain from pushing the box to avoid penalties leading to decreasing overall returns. We verified that this behaviour was consistent even for lower entropy regularisation coefficient $10^{-4}$ for reduced penalty-inducing exploration. We find Ind-MATE and Mix-MATE are significantly more robust at informing agents about the task and allow them to adapt their behaviour by limiting unsuccessful pushing attempts. We hypothesise that Cen-MATE is less successful in these tasks due to its worse sample efficiency exhibiting lower returns at the end of the initial training phase. Agents fine-tuned with MATE are also better at adapting to tasks with larger gridworlds, reaching higher returns in particular for Mix-MATE and Ind-MATE.

\looseness=-1
In LBF (\Cref{fig:finetuning_lbf_all}), we find that the majority of tasks require comparably little adaptation with agents reaching similar returns at the beginning and end of fine-tuning. We hypothesise that the high degree of variability within individual LBF tasks, through agent and food placement as well as level allocation, allows agents to already learn most capabilities required for testing tasks during the initial training phase. In tasks with penalty for unsuccessful picking-up of food, significant adaptation of behaviour is required but policies obtained after the initial training phase tend to frequently attempt to pick-up food. This behaviour results in lower returns and consequentially negative initial transfer performance.

Overall, we find MATE to improve teamwork adaptation in several cases compared to both baselines. In particular in the BPUSH environment where all agents need to coordinate, we find MATE consistently improves returns during fine-tuning. Across all three paradigms of MATE, we find Cen-MATE to perform the best in environments with significant partial observability such as most RWARE tasks. However, training the shared encoder on the larger joint observations of all agents makes training with Cen-MATE less sample efficient leading to worse returns at the end of fine-tuning in several BPUSH tasks and the corridor task with four agents (\Cref{fig:finetuning_rware_corridor_4ag}). Lastly, we find the mixture of task embeddings appears to improve adaptation with Mix-MATE outperforming Ind-MATE in almost all fine-tuning tasks. Due to these benefits, we recommend to learn task embeddings using Mix-MATE whenever decentralised execution is of importance. For testing tasks with significant partial observability, Cen-MATE should be considered for its privileged encoded information but prevents decentralised deployment of agents.

\begin{figure}[t]
    \begin{subfigure}{.02\textwidth}
        \centering
        \includegraphics[angle=90, height=15em]{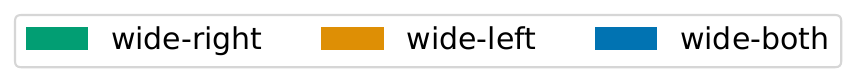}
    \end{subfigure}
    \begin{subfigure}{.40\textwidth}
        \centering
        \includegraphics[width=.8\textwidth]{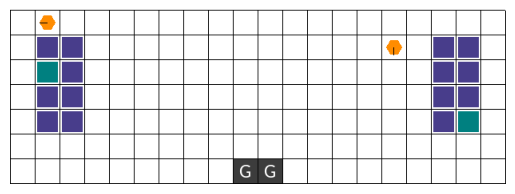}
        \includegraphics[width=.8\textwidth]{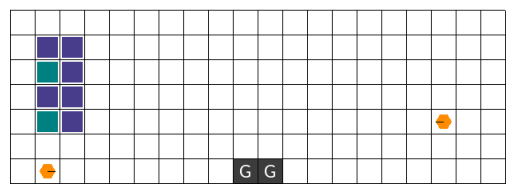}
        \includegraphics[width=.8\textwidth]{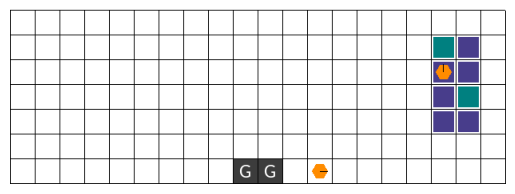}
    \end{subfigure}
    \begin{subfigure}{.5\textwidth}
        \centering
        \includegraphics[trim={0 0 0 3em},clip,width=\textwidth]{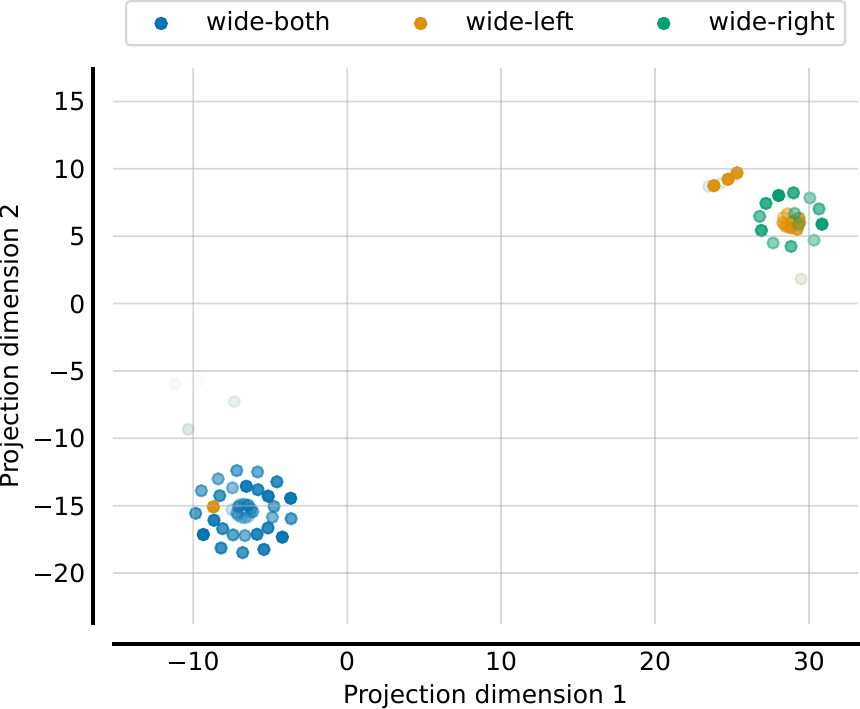}
    \end{subfigure}
    \caption{We collect task embeddings across multiple warehouse environments for a single episode using Cen-MATE trained in one-sided warehouses. We project the task embeddings with t-SNE into two dimensions and visualise embeddings across episodes with colours representing warehouse layouts and opacity indicating time steps within the episodes (more opaque for later time steps).}
    \label{fig:mate_tsne}
\end{figure}

\subsection{Learned Multi-Agent Task Embeddings}
\label{sec:evaluation_analysis_embeddings}

After evaluating the benefits of MATE for teamwork adaptation, we analyse the task embeddings learned by MATE to gain a deeper understanding for the proposed method. First, we address the questing whether task embeddings encode meaningful information to identify tasks. \Cref{fig:mate_tsne} visualises the t-SNE~\citep{van2008visualizing} projection of task embeddings obtained using Cen-MATE for both one-sided wide warehouses as well as the wide warehouse with shelves on both sides. More specifically, we execute a policy trained without MATE in one-sided warehouses and sequentially encode the trajectory using Cen-MATE trained in the same one-sided warehouses. \Cref{fig:mate_tsne} clearly illustrates separate clusters of embeddings for different tasks. Whereas the warehouse with shelves on both sides (blue) is clearly separated from one-sided warehouses, embeddings for both symmetric one-sided warehouses are close together. However, embeddings of one-sided warehouses are also separated with embeddings for the wide-left (orange) warehouse concentrated within a ring of embeddings for the wide-right (green) warehouse. Note that these embeddings were not trained in the wide-both warehouse task, demonstrating that MATE is able to generalise to novel tasks and produce task embeddings which identify tasks not encountered during training.

\begin{figure}[t]

    \centering
    \begin{subfigure}{.50\textwidth}
        \centering
        \includegraphics[width=.95\textwidth]{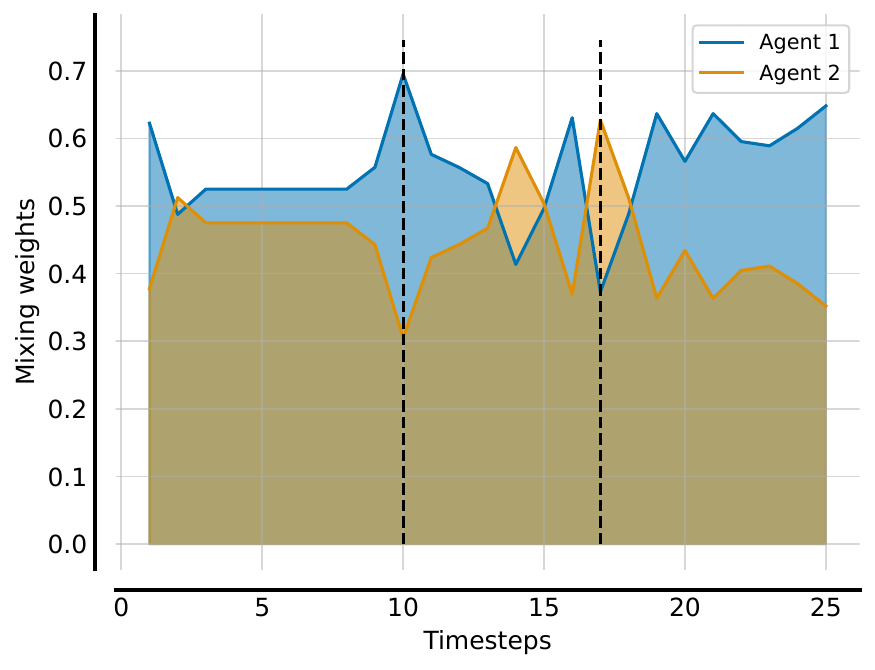}
    \end{subfigure}
    \begin{subfigure}{.45\textwidth}
        \centering
        \includegraphics[width=\textwidth]{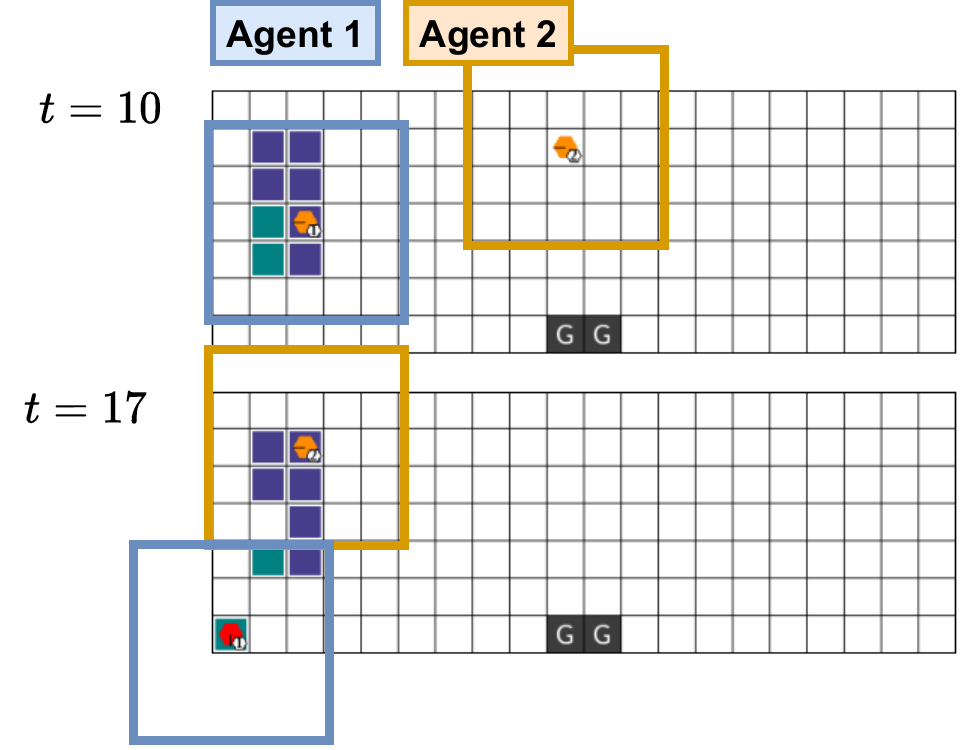}
    \end{subfigure}
    \caption{Visualisation of Mix-MATE mixing weights for 25 time steps in the wide-left warehouse environment (left). At time step 10 (right top), agent 1 is observing the shelves in the warehouse while agent 2 does not observe any shelves so higher weighting is assigned to its respective embedding. At time step 17 (right bottom), agent 2 now observes a larger part of the shelves which results in higher weights on the embedding of agent 2.}
    \label{fig:mate_mixing}
\end{figure}

In \Cref{sec:evaluation}, we find Mix-MATE to consistently outperform Ind-MATE for teamwork adaptation. It appears the mixture model significantly improves the impact of learned task embeddings on teamwork adaptation. \Cref{fig:mate_mixing} visualises mixture weights for Mix-MATE in the wide-left warehouse environment. The mixing has no significant emphasis on the embedding of either agent until the first agent reaches the block of shelves on the left. The discovery and encoding of this task-relevant information leads to a rise in weighting on task embedding of the first agent. This importance ends after the second agent also reaches the block of shelves. After both agents discovered the shelves identifying the task, the mixing appears to again focus on the first agent which has loaded a shelf with requested items.

\section{Conclusion}
In this work, we discussed the problem of teamwork adaptation in which a team of agents needs to adapt their policies to solve novel tasks with limited fine-tuning. We presented three paradigms to learn multi-agent task embeddings (MATE), which recurrently encode information about the current task identified by its unique transition and reward functions. Our experiments in four multi-agent environments demonstrated that MATE provides useful information for teamwork adaptation, with most significant improvements in tasks where all agents need to coordinate. Our analysis of learned task embeddings shows that they form clusters which identify tasks. Among the three paradigms proposed to learn MATE, we find mixed MATE to perform the best across a wide range of testing tasks. Learned mixtures of task embeddings provide interpretable insight into the training of these embeddings, with higher weights being put on agents which currently observe more information about the task.

Future work could aim to relax the assumption that training and testing tasks have identical observation dimensionality and number of agents. This assumption limits the settings in which MATE can be applied, and could be addressed using network architectures leveraging attention~\citep{vaswani2017attention}. Lastly, future work should evaluate the feasibility of learning a mixture of task embeddings and centrally encoding joint information for centralised embeddings in tasks with many agents to investigate the scalability of our proposed approach.

\bibliographystyle{plainnat} 
\bibliography{references}

\clearpage
\appendix

\section{MATE Loss Derivation}
\label{app:mate_loss_derivation}
The standard evidence lower bound (ELBO) to maximise for variational encoder-decoder networks~\citep{kingma2013auto} with additional hyperparameter $\beta$ proposed to control the KL prior regularisation~\citep{higgins2017beta} is given by
\begin{equation}
    \text{ELBO}(\bphi, \bpsi | \btau_{1:t}) = \mathbb{E}_{q\left({\mathbf{z}_t | \btau_{1:t}; \bphi}\right)}\left[\log p({\mathbf{o}_{t+1}, \mathbf{r}_t | \mathbf{o}_t, \mathbf{a}_t, \mathbf{z}; \bpsi})\right] - \beta \text{KL}\left(q\left({\mathbf{z} | \btau_{1:t}; \bphi}\right)|| p(\mathbf{z})\right)
\end{equation}

For a standard Gaussian prior $p(\mathbf{z}) = \cN(\mathbf{z}; \mathbf{0}, \text{diag}(\mathbf{1}))$ and Gaussian variational distribution $q(\mathbf{z} | \tau_{1:t}) = \cN(\mathbf{z}; \bmu, \text{diag}(\bsigma))$ ($\bmu, \text{diag}(\bsigma) \in \mathbb{R}^d$), we can write a closed-form solution for the KL divergence term:
\begin{equation}
    \mathrm{KL}\left(q\left({\mathbf{z} | \tau_{1:t}; \phi}\right)|| p(\mathbf{z})\right) = -\frac{1}{2} \sum_{j=1}^{d}\left(1+\log \left(\sigma_{j}^{2}\right)-\mu_{j}^{2}-\sigma_{j}^{2}\right)
\end{equation}

Next, we will derive that maximising the log-likelihood of a random variable, such as the decoded observations and rewards, of dimensionality $d_p$ under a generative multivariate Gaussian distribution $p(\mathbf{x} | \mathbf{z}) = \cN(\bmu_p, \text{diag}(\bsigma_p))$ with constant diagonal covariance matrix $\text{diag}(\bsigma_p)$ (1) is equivalent to minimising the Euclidean distance between the random variable and the mean of the multivariate generative distribution. Therefore, minimising the Euclidean distance between samples of the observations and rewards and a deterministic output of the decoder is equivalent to maximising the log-likelihood with the generator learning to predict the mean of the modelled variables. We make use of the fact that we can ignore any constants for the maximisation objective (2):
{
    \small
    \begin{align}
        \max \log p(\mathbf{x} | \mathbf{z}) &= \max \log \left[\frac{1}{\sqrt{(2\pi)^{d_p}}|\text{diag}(\bsigma_p)|} \exp\left(-\frac{1}{2}(\mathbf{x} - \bmu_p)^T \text{diag}(\bsigma_p)^{-1} (\mathbf{x} - \bmu_p)\right)\right]&\\
                                             &= \max \log \left[\frac{1}{\sqrt{(2\pi)^{d_p}}|\text{diag}(\bsigma_p)|} \right] + \log \left[\exp\left(-\frac{1}{2}(\mathbf{x} - \bmu_p)^T \text{diag}(\bsigma_p)^{-1} (\mathbf{x} - \bmu_p)\right)\right]&\\
                                           &\stackrel{(1,2)}{=} \max \log \left[\exp\left(-\frac{1}{2}(\mathbf{x} - \bmu_p)^T \text{diag}(\bsigma_p)^{-1} (\mathbf{x} - \bmu_p)\right)\right]&\\
                                           &= \max -\frac{1}{2}(\mathbf{x} - \bmu_p)^T \text{diag}(\bsigma_p)^{-1} (\mathbf{x} - \bmu_p)&\\
                                           &\stackrel{(1,2)}{=}\max -(\mathbf{x} - \bmu_p)^T (\mathbf{x} - \bmu_p)&\\
                                           &= \min \left(\mathbf{x} - \bmu_p\right)^2
    \end{align}
}

Given this derivation (3), we will rewrite the log-likelihood term in the ELBO assuming that the generator, or decoder, is parameterised as a multivariate Gaussian model over observations $p(\mathbf{o}_{t+1} | \mathbf{o}_t, \mathbf{a}_t, \mathbf{z}) = \cN(\bmu_o, \text{diag}(\bsigma_o))$ and rewards $p(\mathbf{r}_t | \mathbf{o}_t, \mathbf{a}_t, \mathbf{z}) = \cN(\bmu_r, \text{diag}(\bsigma_r))$ with constant diagonal covariance matrices $\text{diag}(\bsigma_o)$ and $\text{diag}(\bsigma_r)$, respectively. Following the assumption of a constant diagonal covariance matrix, we can assume that the multivariate Gaussian models for observations and rewards are independent of each other (4).
\begin{align}
    \max \log p({\mathbf{o}_{t+1}, \mathbf{r}_t | \mathbf{o}_t, \mathbf{a}_t, \mathbf{z}; \psi}) &= \max \log \left[p({\mathbf{o}_{t+1} | \mathbf{o}_t, \mathbf{a}_t, \mathbf{z}; \psi}) \cdot p({\mathbf{o}_{t+1} | \mathbf{o}_t, \mathbf{a}_t, \mathbf{z}; \psi})\right]\\
   &\stackrel{(4)}{=} \max \log p({\mathbf{o}_{t+1} | \mathbf{o}_t, \mathbf{a}_t, \mathbf{z}; \psi}) + \log p({\mathbf{o}_{t+1} | \mathbf{o}_t, \mathbf{a}_t, \mathbf{z}; \psi})\\
   &\stackrel{(3)}{=} \min \left(\mathbf{o}_{t+1} - \bmu_o\right)^2 + \left(\mathbf{r}_t - \bmu_r\right)^2
\end{align}

Following the derivation of the KL regularisation term as well as showing the equivalence of maximising the log-likelihood of observations and rewards to minimising the Euclidean distance between generative Gaussian distribution mean and samples of observations and rewards, we obtain the MATE loss presented in \Cref{eq:mate_loss}:
\begin{multline}
    \mathbb{L}(\bphi, \bpsi | \btau_{1:t})  = \mathbb{E}_{q\left({\mathbf{z}_t | \btau_{1:t}; \bphi}\right)}\Big[\left(p({\mathbf{o}_{t+1} | \mathbf{o}_t, \mathbf{a}_t, \mathbf{z}; \bpsi}) - \mathbf{o}_{t+1}\right)^2 \\+ \left(p({\mathbf{r}_t | \mathbf{o}_t, \mathbf{a}_t, \mathbf{z}; \bpsi}) - \mathbf{r}_t\right)^2\Big] - \beta \frac{1}{2} \sum_{j=1}^{d}\left(1+\log \left(\sigma_{j}^{2}\right)-\mu_{j}^{2}-\sigma_{j}^{2}\right)
\end{multline}

\section{Multi-Agent Environments}
\label{app:marl_envs}
\begin{figure}[t]
    \begin{subfigure}{.39\textwidth}
        \centering
        \includegraphics[width=.8\textwidth]{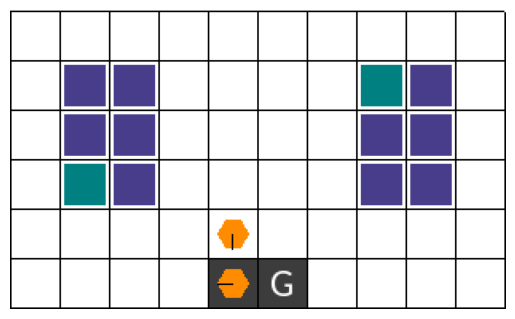}
        \includegraphics[width=.8\textwidth]{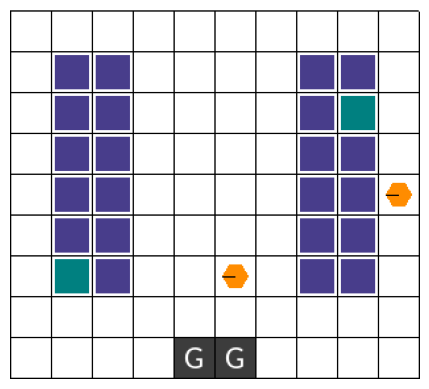}
        \includegraphics[width=.8\textwidth]{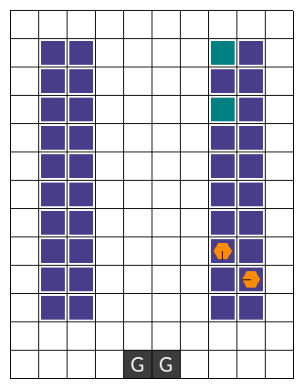}
        \caption{Tiny}
    \end{subfigure}
    \begin{subfigure}{.39\textwidth}
        \centering
        \includegraphics[width=.8\textwidth]{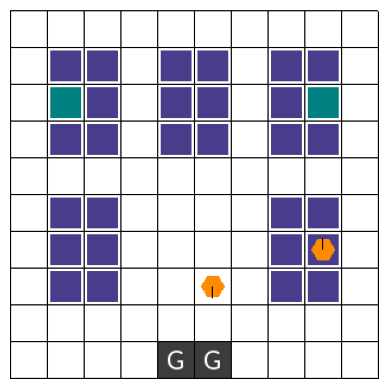}
        \includegraphics[width=.8\textwidth]{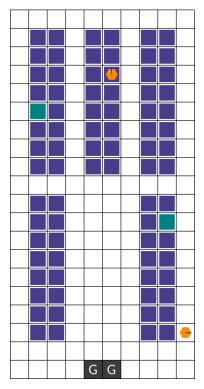}
        \caption{Small}
    \end{subfigure}
    \begin{subfigure}{.20\textwidth}
        \centering
        \includegraphics[width=.8\textwidth]{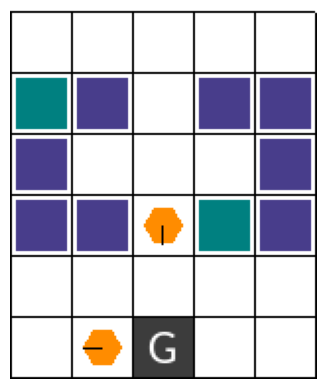}
        \includegraphics[width=.8\textwidth]{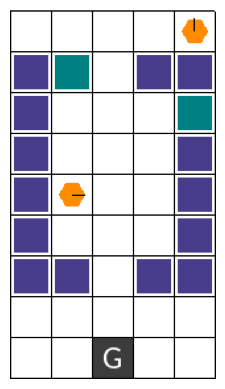}
        \includegraphics[width=.8\textwidth]{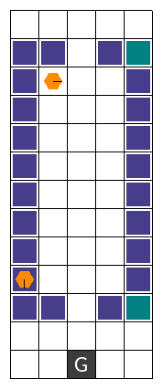}
        \caption{Corridor}
    \end{subfigure}
    \caption{Tiny, small and corridor warehouses with 2 agents and column heights of 3, 6, 10 for tiny and corridor and colum heights 3 and 8 for small.}
    \label{fig:rware_main}
\end{figure}

\subsection{Multi-Robot Warehouse}
The multi-robot warehouse (RWARE)~\citep{papoudakis2021benchmarking,christianos2020shared} environment\footnote{RWARE is licensed under the MIT license.} simulates a logistics task in which agents control robots needing to collect and deliver requested items from shelves within warehouses. The number of requested items at each time step is identical to the number of agents $N$. For this work, we use a modified observation space with a $5\times5$ grid centred around the agent containing information about shelves, requested items, agents, delivery zones and boundaries of the warehouse. All information is encoded as a binary features for $5\times5\times5$ values. Additionally, each agent observes its own rotation within the warehouse as a onehot vector and two binary features indicating if the agent currently carries a shelf and is located on shelves, respectively. All this information is flattened to observations with $131$ values. This observation space was chosen to simplify generalisation by encoding all information relative to the agent~\citep{schaefer2022task} which is further ensured by rotating observations whenever the agent rotates within the warehouse. Agents can choose between five actions: stay, move forward, rotate left, rotate right and toggle load. The last action corresponds to picking-up as well as putting-down shelves depending on the current load of the agent. Agents receive individual rewards of $+1$ for successfully delivering a shelf with a requested item to the delivery zone and $0$ otherwise. Upon successful delivery of a requested item, a new, currently unrequested item is randomly sampled and added to the list of requests to ensure that the number of total requests at each time step is constant and equal to the number of agents $N$.

\begin{figure}[t]
    \begin{subfigure}{.49\textwidth}
        \centering
        \includegraphics[width=\textwidth]{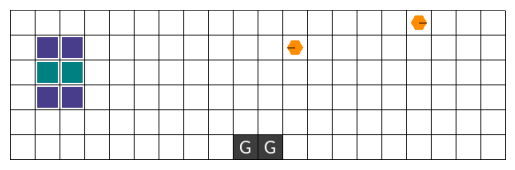}
        \includegraphics[width=\textwidth]{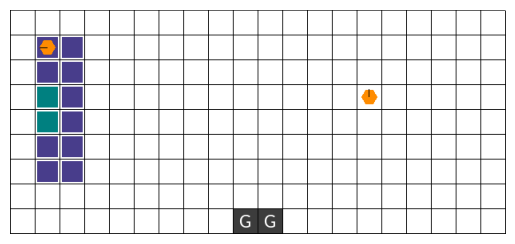}
        \caption{Wide-one-sided left}
    \end{subfigure}
    \begin{subfigure}{.49\textwidth}
        \centering
        \includegraphics[width=\textwidth]{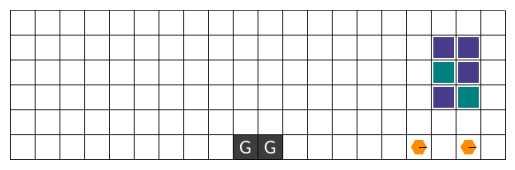}
        \includegraphics[width=\textwidth]{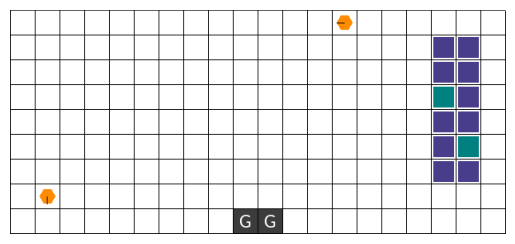}
        \caption{Wide-one-sided right}
    \end{subfigure}
    \begin{subfigure}{.49\textwidth}
        \centering
        \includegraphics[width=\textwidth]{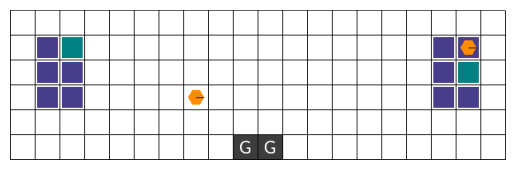}
        \caption{Wide-both}
    \end{subfigure}
    \begin{subfigure}{.49\textwidth}
        \centering
        \includegraphics[width=\textwidth]{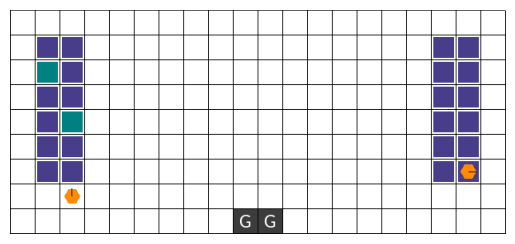}
        \caption{Wide-both}
    \end{subfigure}
    \caption{Wide-one-sided and wide-both warehouses with 2 agents and column heights of 3 and 6.}
    \label{fig:rware_wide}
\end{figure}

In this work, we consider tasks with varying warehouses with two or four agents. Each set of training and testing tasks contains multiple warehouse tasks with varying heights of blocks of shelves between $3$ and $10$. For visualisations of all warehouse layouts, see \Cref{fig:rware_main,fig:rware_wide}.

\subsection{Multi-Agent Particle Environment}
The multi-agent particle environment\footnote{MPE environment is licensed under the MIT license.} (MPE)~\citep{mordatch2018emergence,lowe2017multi} contains several cooperative and competitive multi-agent tasks. Agents control particles in a two-dimensional continuous environment to move with actions being: stay, move right, move left, move up, and move  down. In this work, we consider the cooperative navigation task in which three agents need to move to three landmarks within the environment and avoid collisions with other agents. For each episode, landmarks are randomly placed in the environment but landmarks remain fixed throughout the episode. Agents observe their velocity, position as well as their relative positions to other agents and all landmarks. All agents receive the same rewards defined as the sum of negative minimal distances from each landmark to any agent and an additional penalty reward of $-1$ for collisions of agents. Episodes end after $25$ time steps.

Based on this environment, we further define two new tasks as testing tasks in which agents receive larger negative rewards for collisions with $-5$ and $-50$ penalty rewards for each collision.

\subsection{Boulder-Push}
The Boulder-push environment (BPUSH) is a new environment proposed as part of this work. In this environment, agents navigate a gridworld and are tasked to push a box in a fixed direction towards a given goal location. At the beginning of each episode, agents and the box are randomly placed in the environment and the direction and distance in which the box needs to be pushed are randomly determined. Agents observe a flattened $9\times9$ grid centred on themselves containing the locations of agents and the box as binary features. Additionally, agents globally observe the direction in which the box has to be pushed. Agent actions include movement in each cardinal direction and agents receive a large positive reward of $+1$ for pushing the box to its goal location. Agents also receive a smaller reward of $+0.1$ for pushing the box forward at all. This additional reward is intended to simplify the exploration of agents to discover the optimal policy. The box is only successfully pushed if all agents walk against the box from the opposite side and the box can only be moved in its pre-determined direction. Episodes terminate after the box has reached its goal location or after at most $50$ time steps.

\begin{figure}[t]
    \begin{subfigure}{.33\textwidth}
        \centering
        \includegraphics[width=\textwidth]{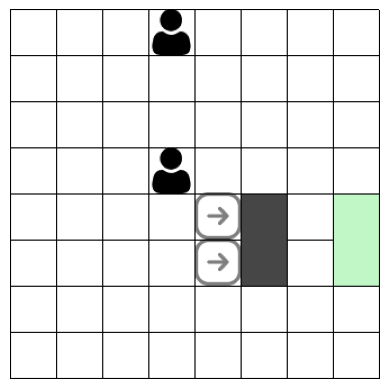}
        \caption{Small}
    \end{subfigure}
    \begin{subfigure}{.33\textwidth}
        \centering
        \includegraphics[width=\textwidth]{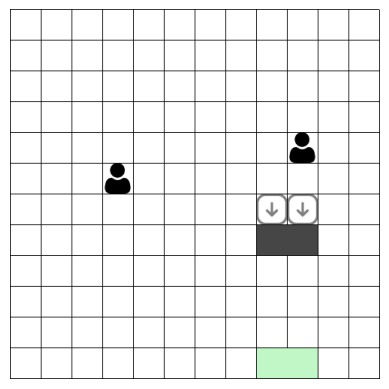}
        \caption{Medium}
    \end{subfigure}
    \begin{subfigure}{.33\textwidth}
        \centering
        \includegraphics[width=\textwidth]{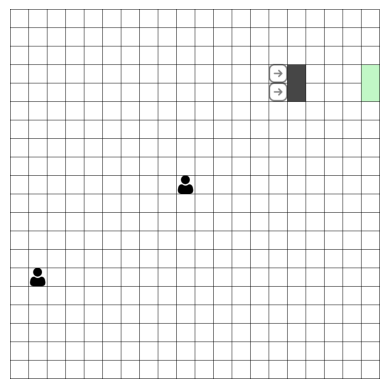}
        \caption{Large}
    \end{subfigure}
    \caption{BPUSH environments with varying sizes and two agents.}
    \label{fig:bpush_renders}
\end{figure}

We initially train two agents in a task with a small $8\times8$ gridworld. Testing tasks include tasks with larger gridworld sizes medium ($12\times12$) and large ($20\times20$) and tasks in which agents receive a small negative reward of $-0.01$ for unsuccessful pushing attempts. Latter occurs whenever a single agent attempts to push the box but the other agent is not cooperating such that the box is not moved.
For visualisations of all considered gridworld sizes with two agents, see \Cref{fig:bpush_renders}.

\subsection{Level-Based Foraging}
The level-based foraging (LBF)~\citep{Albrecht2013ASystems,albrecht2018autonomous} environment\footnote{LBF is licensed under the MIT license.} contains tasks in which agents need to navigate a gridworld to collect food. For each episode, food and agents are randomly placed in the gridworld and assigned levels. Agents are only able to pick-up adjacent food items if the sum of the levels of all agents currently picking-up the food is greater or equal to the level of the food. Each agent observes a $5\times5$ grid centred on itself containing information about nearby agents and food given by their level as well as boundaries of the gridworld. Agents choose one of six discrete actions: stay, move north, move south, move west, move east and pick-up. Agents receive normalised rewards for successful picking-up of food equal to the level of the collected food. It should be noted that agents receive individual rewards rather than sharing a single reward across all agents.

\begin{figure}[t]
    \begin{subfigure}{.33\textwidth}
        \centering
        \includegraphics[width=\textwidth]{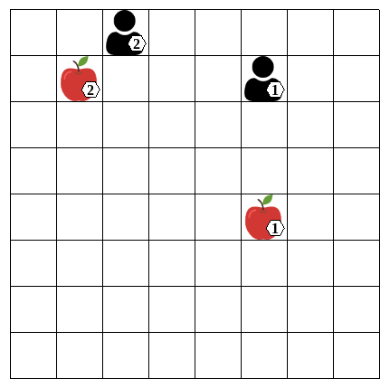}
        \caption{8x8-2p-2f}
    \end{subfigure}
    \begin{subfigure}{.33\textwidth}
        \centering
        \includegraphics[width=\textwidth]{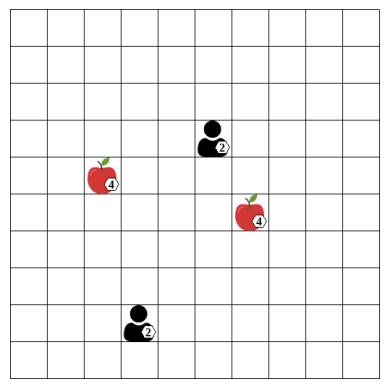}
        \caption{10x10-2p-2f-coop}
    \end{subfigure}
    \begin{subfigure}{.33\textwidth}
        \centering
        \includegraphics[width=\textwidth]{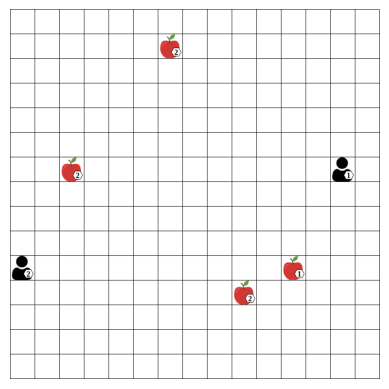}
        \caption{15x15-2p-4f}
    \end{subfigure}
    \caption{LBF tasks with varying sizes and two agents.}
    \label{fig:lbf_renders}
\end{figure}

We initially train two or four agents in tasks with comparably small gridworlds with sizes $8\times8$ containing as many food items as agents. Agent and food levels are randomly assigned for each episode while ensuring that each food can be collected. Therefore, agents will need to cooperate to pick-up some food while others can be collected by individual agents. Testing tasks contain tasks with larger gridworld environments, more food as well as tasks with enforced cooperation or penalty. In tasks with enforced cooperation (denoted by "-coop"), each food has a high level such that it can only be collected if all agents in the environment cooperate. In testing tasks with penalty (denoted by "-pen"), agents receive a negative reward of $-0.1$ for unsuccessful picking attempts. A picking attempt is unsuccessful whenever an agent chose the pick-up action but did not succeed at collecting any food.
For visualisations of all considered gridworld sizes with two agents as well as visualisation of a task with forced cooperation, see \Cref{fig:lbf_renders}.

\section{Experimental Details}
\label{app:imp_details}

\subsection{Hyperparameters}
For all algorithms, we train agent policies using the MAA2C algorithm with each agent optimising its own policy and critic. The critic computes a state value function and is conditioned on the joint observations of all agents. Agents are trained in ten synchronous environments running in parallel~\citep{mnih2016asynchronous} with batches of experience from all environments across 5 time steps used to compute policy, value and MATE loss and optimise the networks. We compute 5-step return estimates for each of the environments and a entropy regularisation term was added to the loss to incentivise exploration~\citep{mnih2016asynchronous,ziebart2010modeling}. Policy networks consist of a single fully-connected layer followed by a gated recurrent unit (GRU)~\citep{cho2014learning} and a last fully-connected layer. Critic networks consist of two fully-connected layers and the ReLU activation function is applied after each layer. Networks are optimised using the Adam~\citep{kingma2014adam} optimiser and target critic networks were used to compute target return estimates and updated using soft updates with $\tau = 0.01$.

MATE encoders consist of a fully-connected layer, followed by a GRU and a final fully-connected layer and decoders consist of two fully-connected layers. The encoder output $\cN(\mathbf{z})$ has dimensionality of $6$ with $|\mu| = |\sigma| = 3$. For MATE hyperparameters, a small gridsearch over $\beta$, task embedding size and learning rate was conducted in the RWARE wide fine-tuning scenarios. We also find that backpropagating gradients from the MARL training to the MATE encoders is possible but does not lead to improved adaptation performance so we stop gradients to flow into the MATE encoders from the MARL training.

For all hyperparameters for MAA2C and MATE, see \Cref{tab:hyperparams}.

\begin{table}[ht!]
    \centering
    \caption{Hyperparameters for MAA2C and MATE}
    \label{tab:hyperparams}
    \begin{tabular}{l l r}
        \toprule
        Algorithm & Hyperparameter & Value\\
        \midrule
        \multirow{13}{*}{MAA2C} & Policy network            & FC(128) $\times$ GRU(128) $\times$ FC(128)\\
                                & Critic network            & FC(128) $\times$ FC(128)\\
                                & Activation function       & ReLU\\
                                & Optimiser                 & Adam\\
                                & Learning rate             & $5e^{-4}$\\
                                & Adam epsilon              & $1e^{-3}$\\
                                & Entropy loss coefficient  & $0.01$\\
                                & Value loss coefficient    & $0.5$\\
                                & Target critic $\tau$      & $0.01$\\
                                & $\gamma$                  & 0.99\\
                                & N-step returns            & 5\\
                                & Parallel environments     & 10\\
                                & Max grad norm             & / (no gradient normalisation)\\ \midrule
        \multirow{10}{*}{MATE}   & Encoder $q$ network       & FC(64) $\times$ GRU(64) $\times$ FC(64)\\
                                & Decoder $p$ network       & FC(64) $\times$ FC(64)\\
                                & Mixing $f_m$ network      & FC(64)\\
                                & Activation function       & ReLU\\
                                & Optimiser                 & Adam\\
                                & Learning rate             & $1e^{-4}$\\
                                & Adam epsilon              & $1e^{-3}$\\
                                & $\beta$                   & $0.1$\\
                                & Task embedding size       & 3\\
                                & Max grad norm             & 0.5\\
        \bottomrule
    \end{tabular}
    
\end{table}

\subsection{Computational Resources}
Experiments were run on CPUs only across three research clusters being equipped with (1) a AMD EPYC 7502 CPU with 32 cores @2.5Ghz running Ubuntu 20.04.3 LTS, (2) a AMD EPYC 7502 CPU running Scientific Linux 7.9, and (3) a AMD  EPYC 7H12 CPU with 64 cores @2.6Ghz running Ubuntu 20.04.3 LTS. We used Python 3.9.7 with PyTorch 1.10~\citep{pytorch2019}.

We conduct a runtime comparison of training agents with and without MATE and present the runtime together with percentage increase of using MATE in comparison to using no MATE in two tasks in \Cref{tab:runtime}. For this comparison, we train each of the algorithms for 100,000 time steps in the RWARE tiny-2ag and RWARE tiny-4ag tasks to also indicate the scalability with varying number of agents. Timings were measured on a personal computer with a Intel i7-10750H CPU with 12 cores @2.60GHz running Ubuntu 20.04.4 LTS, Python 3.9.7 and PyTorch 1.10. For all algorithms, we used the hyperparameters specified in \Cref{tab:hyperparams}.

\begin{table}[ht!]
    \centering
    \caption{Runtime (in seconds) for agents trained using MAA2C with and without MATE for 100,000 time steps.}
    \label{tab:runtime}
    \begin{tabular}{l l l l}
        \toprule
        Task & Algorithm & Runtime & \% Increase\\
        \midrule
        \multirow{4}{*}{RWARE tiny-2ag} & MAA2C No MATE     & 144.32s & /\\
                                        & MAA2C Ind-MATE    & 170.02s & 17.81\%\\
                                        & MAA2C Cen-MATE    & 161.06s & 11.60\%\\
                                        & MAA2C Mix-MATE    & 168.38s & 16.67\%\\\midrule
        \multirow{4}{*}{RWARE tiny-4ag} & MAA2C No MATE     & 204.00s & /\\
                                        & MAA2C Ind-MATE    & 243.69s & 19.46\%\\
                                        & MAA2C Cen-MATE    & 223.80s & 9.71\%\\
                                        & MAA2C Mix-MATE    & 250.06s & 22.58\%\\
        \bottomrule
    \end{tabular}
\end{table}

\clearpage
\section{Training Returns}
\label{app:training_plots}

\begin{figure}[ht!]
    \centering
    \includegraphics[trim={0 0.5em 0 0.5em},clip,width=.8\textwidth]{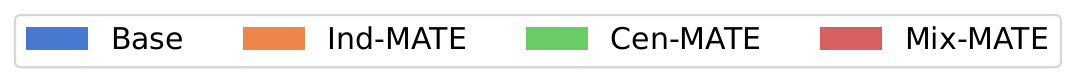}
\end{figure}

\begin{figure}[ht!]
    \centering
    \begin{subfigure}{.45\textwidth}
        \centering
        \includegraphics[width=.9\textwidth]{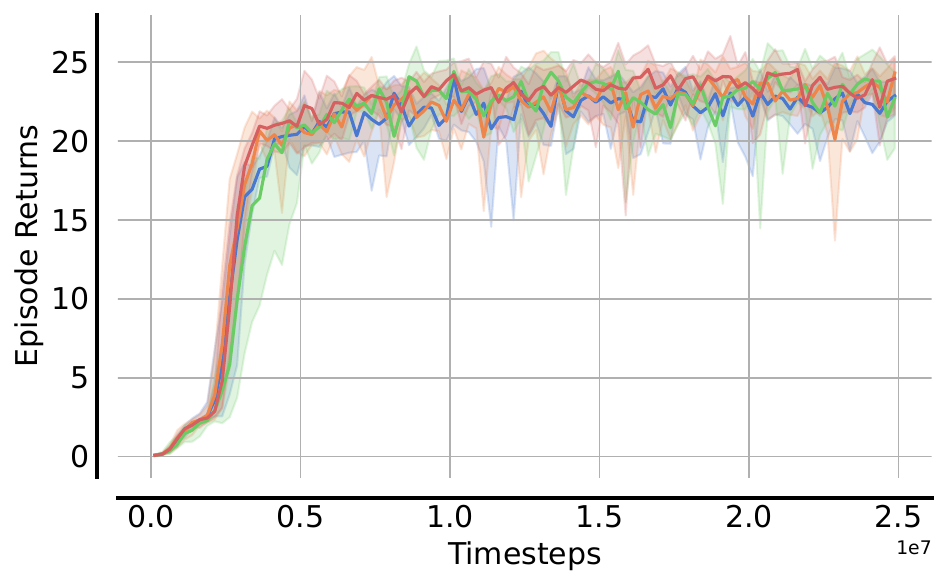}
        \caption{Tiny-2ag}
    \end{subfigure}
    \hspace{2em}
    \begin{subfigure}{.45\textwidth}
        \centering
        \includegraphics[width=.9\textwidth]{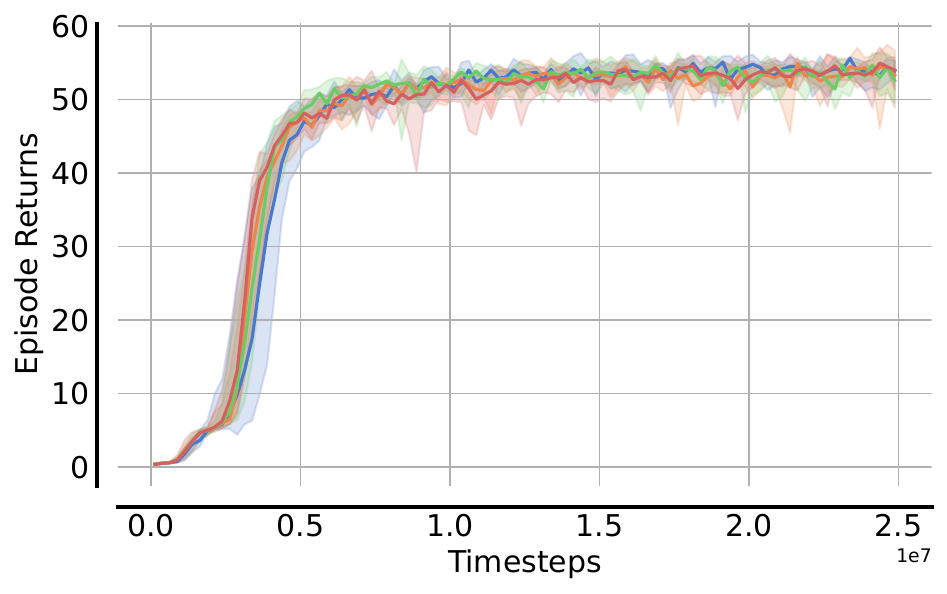}
        \caption{Tiny-4ag}
    \end{subfigure}
    \begin{subfigure}{.45\textwidth}
        \centering
        \includegraphics[width=.9\textwidth]{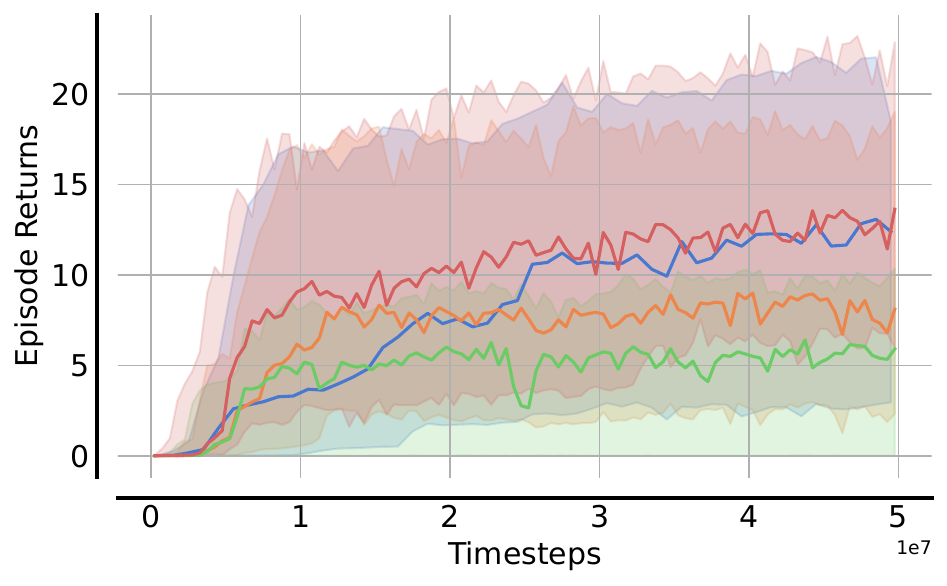}
        \caption{Wide-one-sided}
    \end{subfigure}
    \hspace{2em}
    \begin{subfigure}{.45\textwidth}
        \centering
        \includegraphics[width=.9\textwidth]{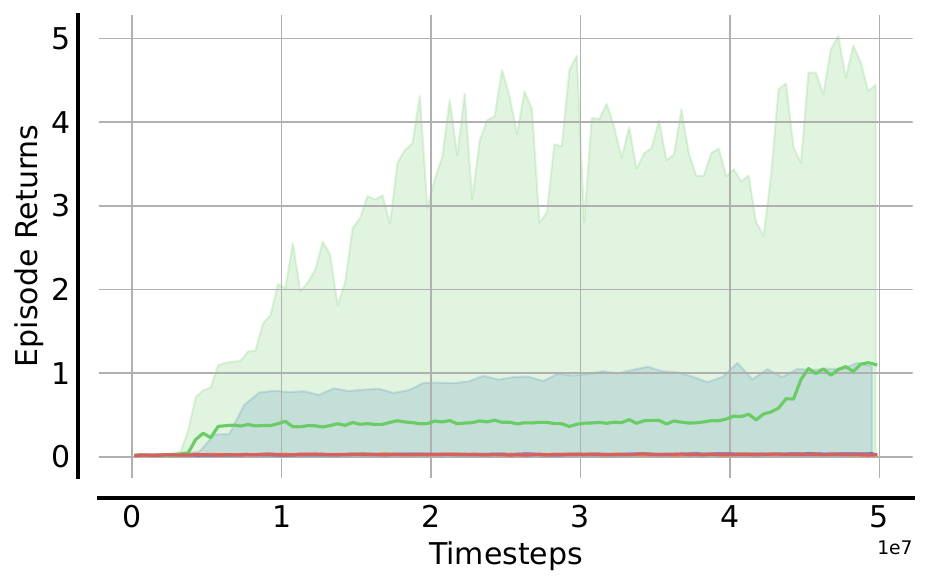}
        \caption{Wide-both}
    \end{subfigure}
    \caption{IQM and 95\% confidence intervals for RWARE training}
    \label{fig:training_rware_all}
\end{figure}

\begin{figure}[ht!]
    \centering
    \begin{subfigure}{.45\textwidth}
        \centering
        \includegraphics[width=.9\textwidth]{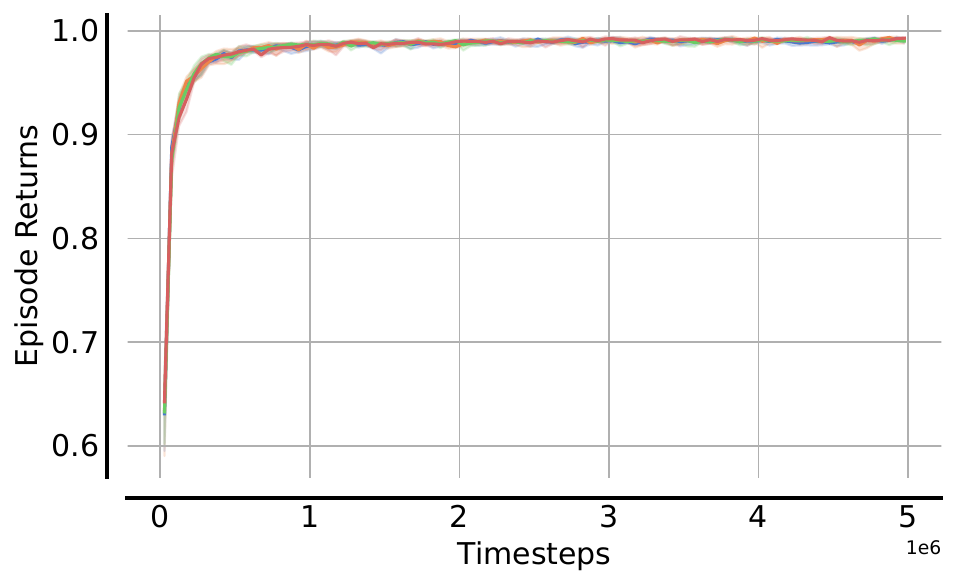}
        \caption{8x8-2p-2f}
    \end{subfigure}
    \hspace{2em}
    \begin{subfigure}{.45\textwidth}
        \centering
        \includegraphics[width=.9\textwidth]{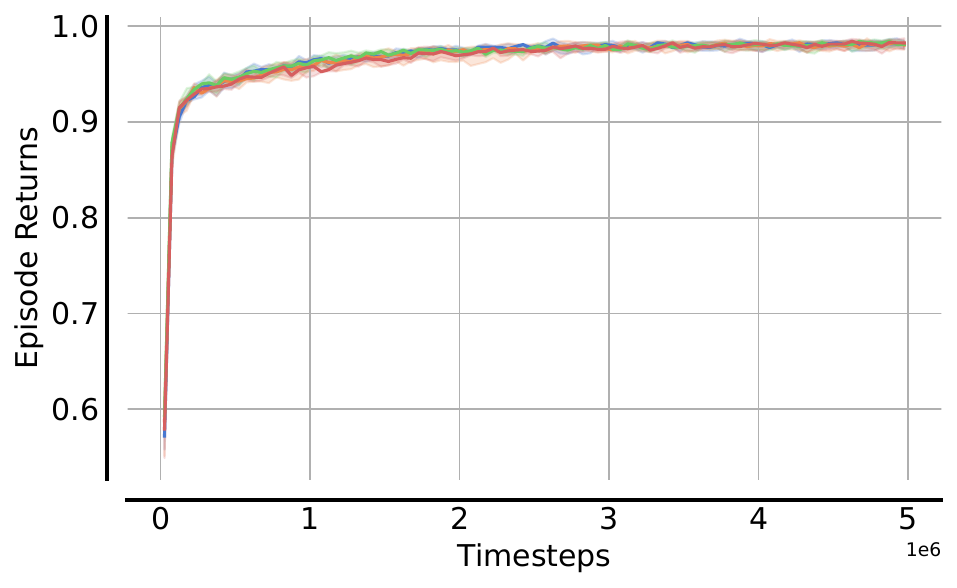}
        \caption{8x8-4p-4f}
    \end{subfigure}
    \caption{IQM and 95\% confidence intervals for LBF training}
    \label{fig:training_lbf_all}
\end{figure}

\begin{figure}[ht!]
    \begin{subfigure}{.45\textwidth}
        \centering
        \includegraphics[width=.9\textwidth]{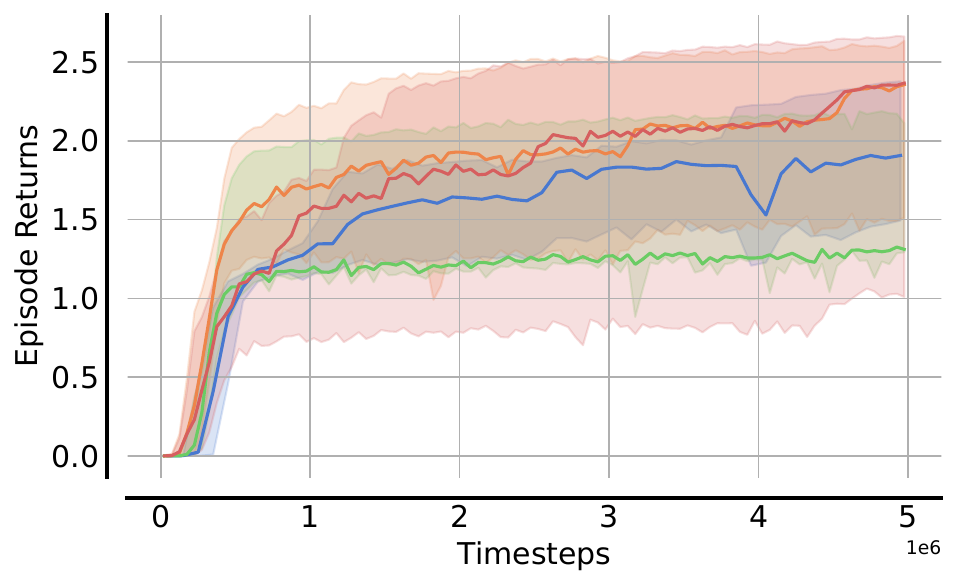}
        \caption{BPUSH small-2ag}
    \end{subfigure}
    \hspace{2em}
    \begin{subfigure}{.45\textwidth}
        \centering
        \includegraphics[width=.9\textwidth]{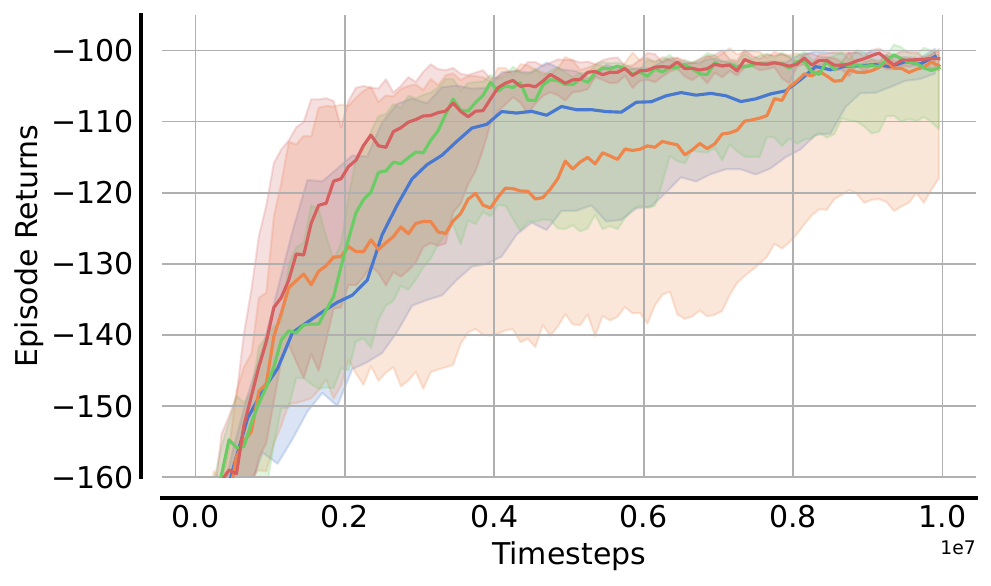}
        \caption{MPE cooperative navigation}
    \end{subfigure}
    \caption{IQM and 95\% confidence intervals for BPUSH and MPE training}
    \label{fig:training_bpush_mpe_all}
\end{figure}

\clearpage
\section{Fine-Tuning Returns}
\label{app:finetuning_plots}

\begin{figure}[ht!]
    \centering
    \includegraphics[trim={0 0.5em 0 0.5em},clip,width=.8\textwidth]{media/results_nolabels/finetuning/legend.pdf}
\end{figure}

\begin{figure}[ht!]
    \begin{subfigure}{.45\textwidth}
        \centering
        \includegraphics[width=.9\textwidth]{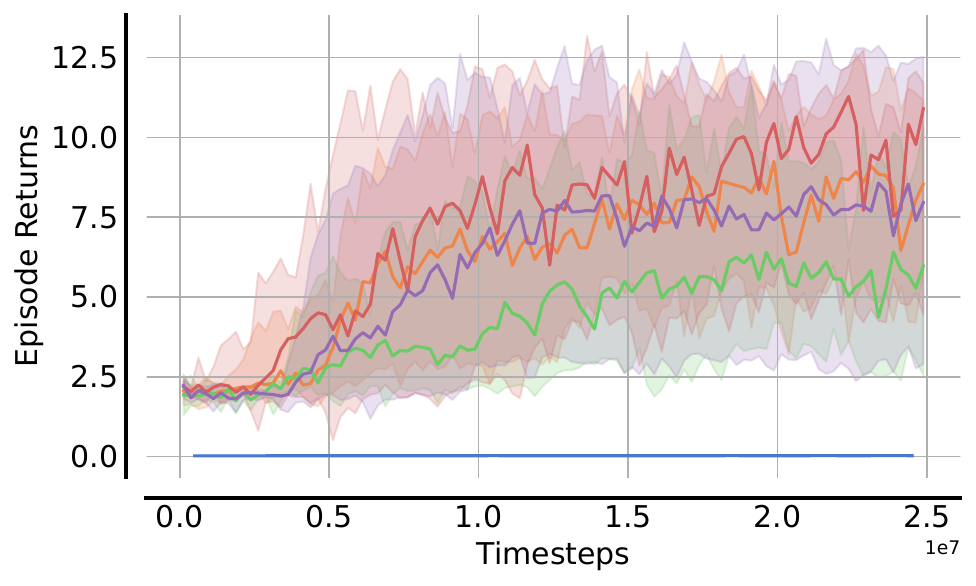}
        \caption{Tiny-2ag $\rightarrow$ small-2ag}
    \end{subfigure}
    \hspace{2em}
    \begin{subfigure}{.45\textwidth}
        \centering
        \includegraphics[width=.9\textwidth]{media/results_nolabels/finetuning/rware/tiny-2ag_corridor-2ag.pdf}
        \caption{Tiny-2ag $\rightarrow$ corridor-2ag}
    \end{subfigure}

    \begin{subfigure}{.45\textwidth}
        \centering
        \includegraphics[width=.9\textwidth]{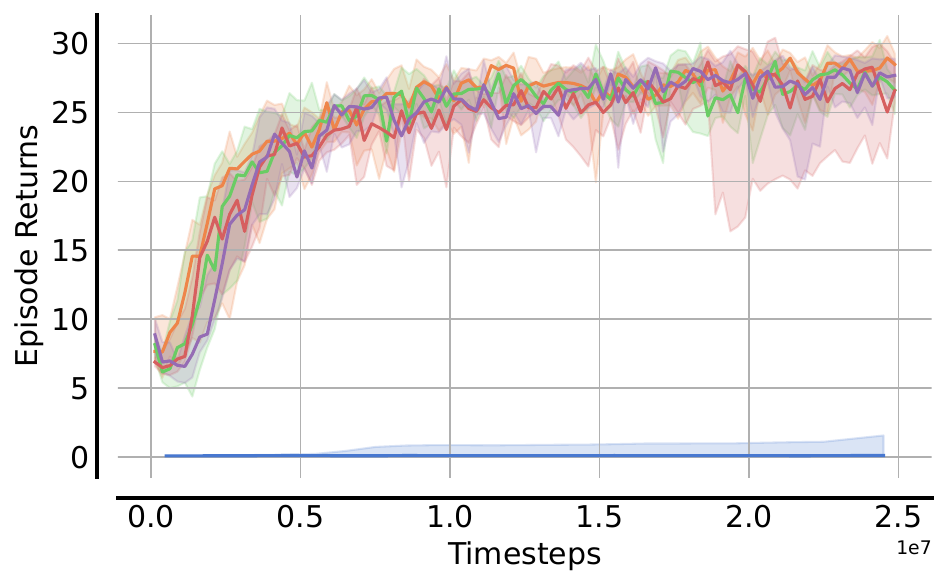}
        \caption{Tiny-4ag $\rightarrow$ small-4ag}
    \end{subfigure}
    \hspace{2em}
    \begin{subfigure}{.45\textwidth}
        \centering
        \includegraphics[width=.9\textwidth]{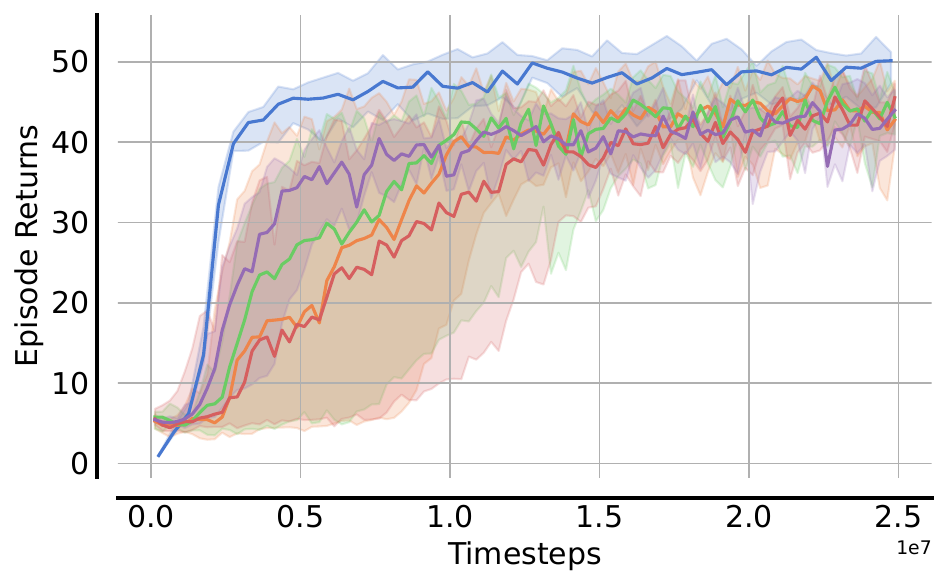}
        \caption{Tiny-4ag $\rightarrow$ corridor-4ag}
        \label{fig:finetuning_rware_corridor_4ag}
    \end{subfigure}

    \begin{subfigure}{.45\textwidth}
        \centering
        \includegraphics[width=.9\textwidth]{media/results_nolabels/finetuning/rware/wide-sides_wide.pdf}
        \caption{Wide-one-sided $\rightarrow$ wide-both}
    \end{subfigure}
    \hspace{2em}
    \begin{subfigure}{.45\textwidth}
        \centering
        \includegraphics[width=.9\textwidth]{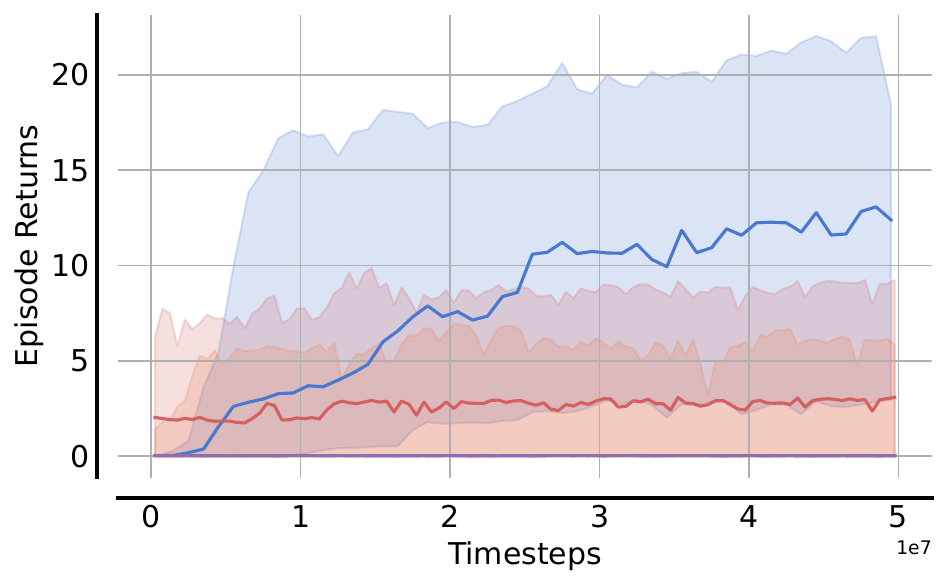}
        \caption{Wide-both $\rightarrow$ wide-one-sided}
    \end{subfigure}
    \caption{IQM and 95\% confidence intervals for RWARE fine-tuning}
    \label{fig:finetuning_rware_all}
\end{figure}

\clearpage
\begin{figure}[ht!]
    \centering
    \includegraphics[trim={0 0.5em 0 0.5em},clip,width=.8\textwidth]{media/results_nolabels/finetuning/legend.pdf}
\end{figure}

\begin{figure}[ht!]
    \begin{subfigure}{.45\textwidth}
        \centering
        \includegraphics[width=.9\textwidth]{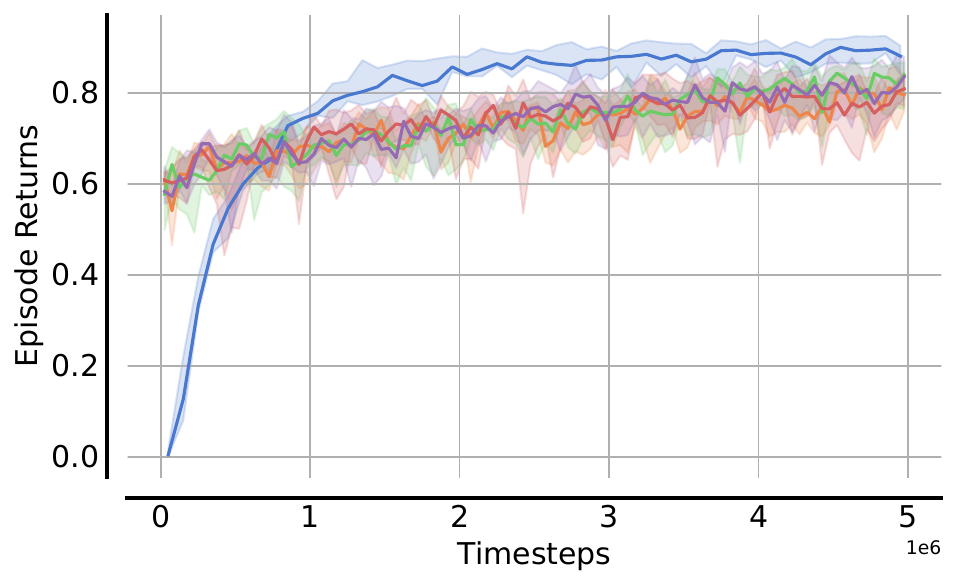}
        \caption{8x8-2p-2f $\rightarrow$ 10x10-2p-2f-coop}
    \end{subfigure}
    \hspace{2em}
    \begin{subfigure}{.45\textwidth}
        \centering
        \includegraphics[width=.9\textwidth]{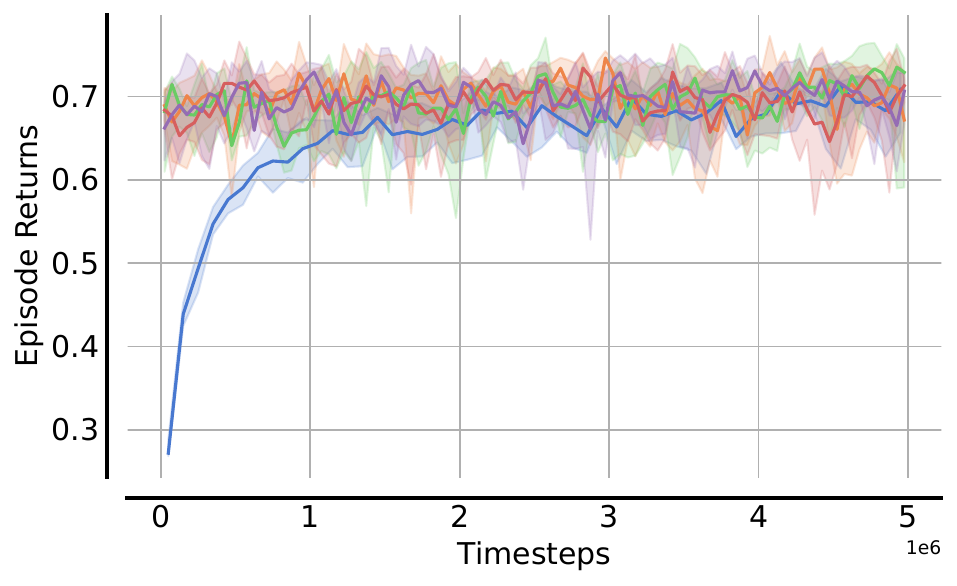}
        \caption{8x8-2p-2f $\rightarrow$ 15x15-2p-4f}
    \end{subfigure}

    \begin{subfigure}{.45\textwidth}
        \centering
        \includegraphics[width=.9\textwidth]{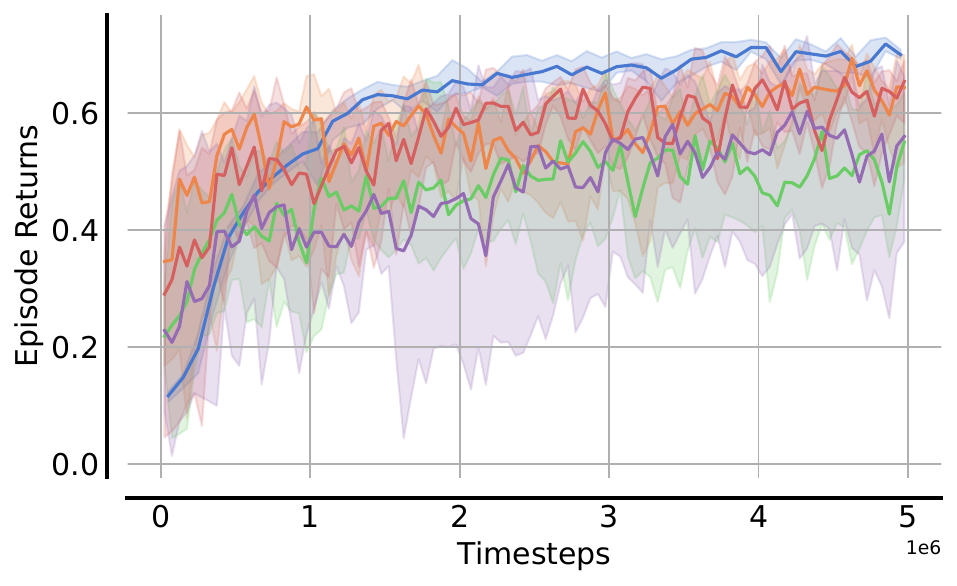}
        \caption{8x8-2p-2f $\rightarrow$ 15x15-2p-2f-pen}
    \end{subfigure}
    \hspace{2em}
    \begin{subfigure}{.45\textwidth}
        \centering
        \includegraphics[width=.9\textwidth]{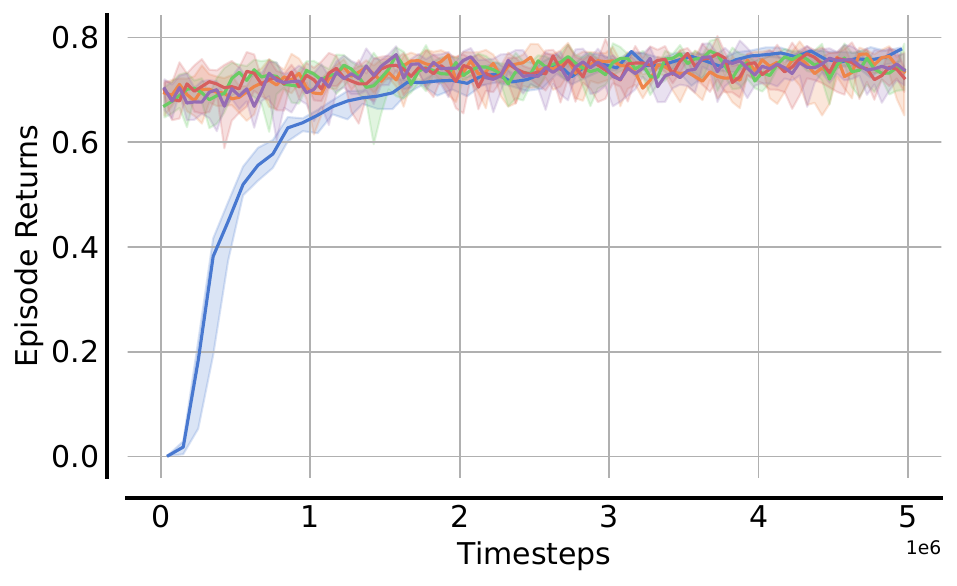}
        \caption{8x8-4p-4f $\rightarrow$ 10x10-4p-2f-coop}
    \end{subfigure}

    \begin{subfigure}{.45\textwidth}
        \centering
        \includegraphics[width=.9\textwidth]{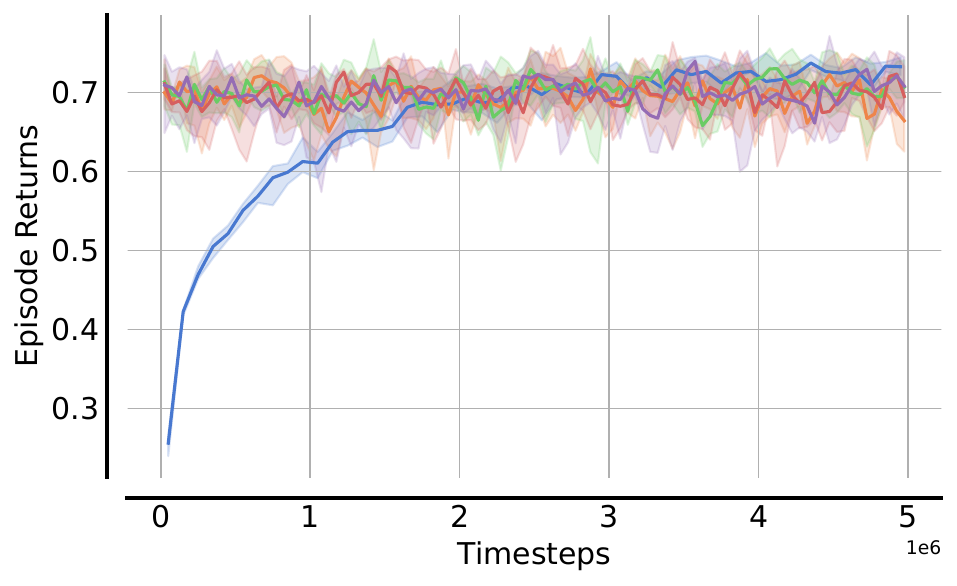}
        \caption{8x8-4p-4f $\rightarrow$ 15x15-4p-6f}
    \end{subfigure}
    \hspace{2em}
    \begin{subfigure}{.45\textwidth}
        \centering
        \includegraphics[width=.9\textwidth]{media/results_nolabels/finetuning/lbf/8x8-4p-4f_15x15-4p-4f-pen.pdf}
        \caption{8x8-4p-4f $\rightarrow$ 15x15-4p-4f-pen}
    \end{subfigure}
    \caption{IQM and 95\% confidence intervals for LBF fine-tuning}
    \label{fig:finetuning_lbf_all}
\end{figure}

\clearpage
\begin{figure}[ht!]
    \centering
    \includegraphics[trim={0 0.5em 0 0.5em},clip,width=.8\textwidth]{media/results_nolabels/finetuning/legend.pdf}
\end{figure}

\begin{figure}[ht!]
    \begin{subfigure}{.45\textwidth}
        \centering
        \includegraphics[width=.9\textwidth]{media/results_nolabels/finetuning/bpush/small-2ag_small-pen-2ag.pdf}
        \caption{Small-2ag $\rightarrow$ small-pen-2ag}
    \end{subfigure}
    \hspace{2em}
    \begin{subfigure}{.45\textwidth}
        \centering
        \includegraphics[width=.9\textwidth]{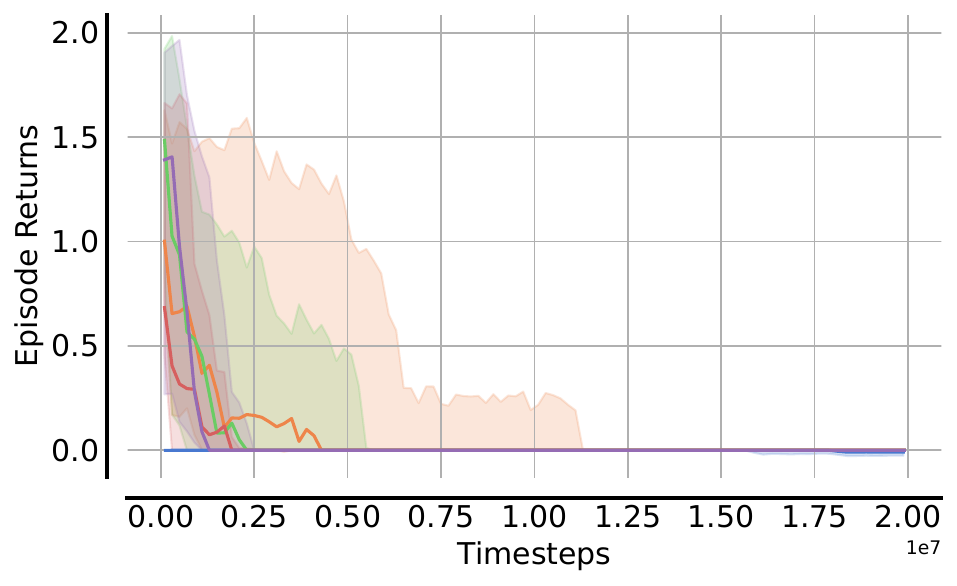}
        \caption{Small-2ag $\rightarrow$ medium-pen-2ag}
    \end{subfigure}

    \begin{subfigure}{.45\textwidth}
        \centering
        \includegraphics[width=.9\textwidth]{media/results_nolabels/finetuning/bpush/small-2ag_medium-2ag.pdf}
        \caption{Small-2ag $\rightarrow$ medium-2ag}
    \end{subfigure}
    \hspace{2em}
    \begin{subfigure}{.45\textwidth}
        \centering
        \includegraphics[width=.9\textwidth]{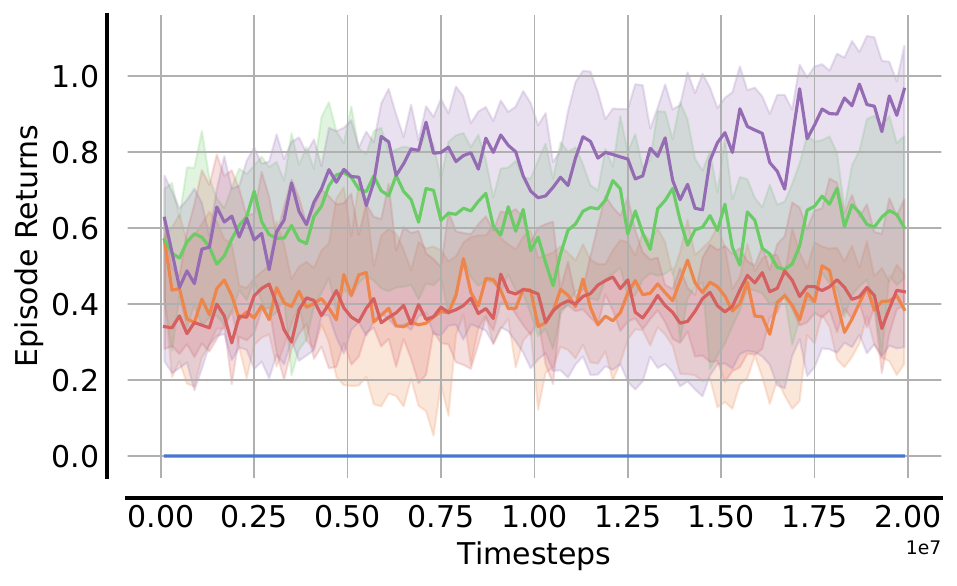}
        \caption{Small-2ag $\rightarrow$ large-2ag}
    \end{subfigure}
    \caption{IQM and 95\% confidence intervals for BPUSH fine-tuning}
    \label{fig:finetuning_bpush_all}
\end{figure}

\begin{figure}[ht!]
    \begin{subfigure}{.45\textwidth}
        \centering
        \includegraphics[width=.9\textwidth]{media/results_nolabels/finetuning/mpe/zone-spread-v4_penalty-spread-v1.pdf}
        \caption{Cooperative navigation (pen=1) $\rightarrow$ penalty navigation (pen=5)}
    \end{subfigure}
    \hspace{2em}
    \begin{subfigure}{.45\textwidth}
        \centering
        \includegraphics[width=.9\textwidth]{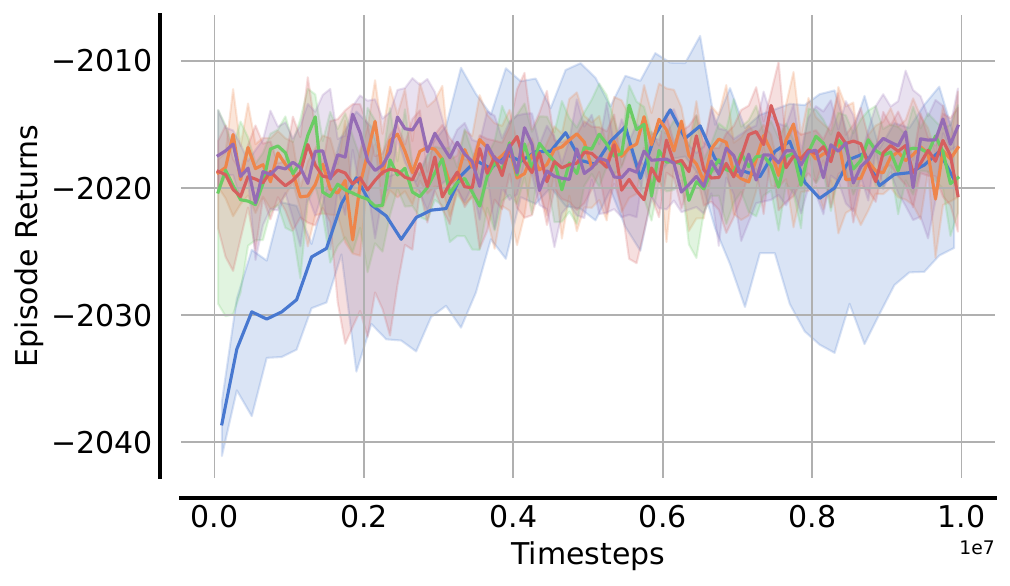}
        \caption{Cooperative navigation $\rightarrow$ penalty navigation (pen=50)}
    \end{subfigure}
    \caption{IQM and 95\% confidence intervals for MPE fine-tuning}
    \label{fig:finetuning_mpe_all}
\end{figure}

\end{document}